\documentstyle[preprint,aps,graphicx,floats]{revtex}

\begin{document}

\preprint{$
\begin{array}{l}
\mbox{UB-HET-06-02}\\[-3mm]
\mbox{November~2006} \\ [3mm]
\end{array}
$}

\title{Weak Boson Emission in Hadron Collider Processes\\
\phantom{trick}}

\author{U.~Baur\footnote{e-mail: baur@ubhex.physics.buffalo.edu}}
\address{Department of Physics,
State University of New York, \\
Buffalo, NY 14260, USA}

\maketitle 

\begin{abstract}
\baselineskip13.pt  
The ${\cal O}(\alpha)$ virtual weak radiative corrections to many
hadron collider 
processes are known to become large and negative at high energies, due to the
appearance of Sudakov-like logarithms. At the same order in perturbation
theory, weak boson emission diagrams contribute. Since the $W$ and $Z$
bosons are massive, the ${\cal O}(\alpha)$ virtual weak radiative
corrections and the 
contributions from weak boson emission are separately finite. Thus,
unlike in QED or QCD calculations, there is no technical reason for
including gauge boson emission diagrams in calculations of electroweak
radiative corrections. In
most calculations of the ${\cal O}(\alpha)$ electroweak radiative
corrections, weak boson emission diagrams are therefore not taken into
account. Another 
reason for not including these diagrams is that they lead
to final states which differ from that of the original process. However,
in experiment, one usually 
considers partially inclusive final states. Weak boson emission diagrams
thus should be included in calculations of 
electroweak radiative corrections. In this paper, I examine the role
of weak boson emission in those processes at the Fermilab
Tevatron and the 
CERN LHC for which the one-loop electroweak radiative corrections are
known to become large at high energies (inclusive jet, isolated photon,
$Z+1$~jet, Drell-Yan, di-boson, $\bar tt$, and 
single top production). In general, I find that the cross section for
weak boson emission 
is substantial at high energies and that weak boson emission and the
${\cal O}(\alpha)$ virtual weak radiative corrections partially cancel. 
\end{abstract}

\newpage


\tightenlines

\section{Introduction}
In 2007, the Large Hadron Collider (LHC) at CERN will begin
operation. One of the main goals of the LHC experiments is to find the
Higgs boson, or, more generally to unveil the mechanism of electroweak
symmetry breaking. In order to discover the Higgs boson, or new physics,
it is necessary to fully understand Standard Model (SM) processes. With
the precision expected from LHC experiments, the theoretical
uncertainties for many processes have to be reduced to the few percent 
level. In addition to the next-to-leading (NLO) and, in some cases,
next-to-next-to-leading (NNLO) QCD corrections, this requires knowledge
of the ${\cal O}(\alpha)$ electroweak (EW) radiative corrections.

The ${\cal O}(\alpha)$ virtual weak radiative corrections are known to
become large and negative at high energies, due to the presence of
Sudakov-like logarithms of the form $(\alpha/\pi)\log^2(\hat
s/M_{W,Z}^2)$, where $\hat s$ is the squared parton center of mass
energy, and $M_{W,Z}$ is the mass of the $W$ or $Z$ boson. For
$\sqrt{\hat s}\geq 1$~TeV, the ${\cal O}(\alpha)$ one-loop EW radiative
corrections can easily become larger in magnitude than the ${\cal
O}(\alpha_s)$ QCD corrections. 

The source of the Sudakov-like logarithms is well
understood~\cite{Ciafaloni:1998xg,Ciafaloni:2000df,Ciafaloni:2000gm,Denner:2000jv,Kuhn:1999de,Melles:2001ye}.
They originate from collinear and infrared divergences which would be
present in the limit of vanishing $W$ and $Z$ masses. In QED, these
divergences are canceled by the corresponding divergences which originate
from real photon radiation diagrams~\cite{kln}. Because the
underlying gauge symmetry is Abelian, observables which include soft
photons are infrared safe (Bloch-Nordsieck theorem)~\cite{bn}. In
non-abelian gauge theories, the Bloch-Nordsieck theorem is violated. This
was initially pointed out for QCD~\cite{bnviol}. However, the infrared
divergences present at the parton level in QCD have no practical consequences
since one sums or averages over color charges when calculating physical
observables. For electroweak interactions this is not automatically the
case and large Sudakov-like logarithms may survive. 

The appearance of large logarithms in one-loop weak corrections has been
demonstrated in a number of explicit calculations. For hadron colliders,
the ${\cal O}(\alpha)$ virtual weak corrections to inclusive
jet~\cite{Moretti:2005ut}, isolated photon~\cite{Kuhn:2005gv,Maina:2004rb},
$Z+1$~jet~\cite{Maina:2004rb,Kuhn:2004em},
Drell-Yan~\cite{Baur:2001ze,kramer,Baur:2004ig,Arbuzov:2005dd,Zykunov:2005tc},
di-boson~\cite{Accomando:2001fn,wgam,Accomando:2004de}, $\bar
tt$~\cite{Kuhn:2005it,mnr,Beenakker:1993yr,Bernreuther:2006vg}, and single top
production~\cite{Beccaria:2006ir,Beccaria:2006dt,Ciafaloni:2006qu} have
been calculated. In almost all calculations, weak boson emission
diagrams have not been taken into account, although
they contribute at the same order in perturbation theory as the one-loop
corrections. From the theoretical point of view this is justified. Since
the $W$ and $Z$ masses act as infrared regulators, the weak boson
emission diagrams are not necessary to obtain a finite answer (in
contrast to QED or QCD corrections). Furthermore, since $W$ and $Z$
bosons decay, weak boson emission diagrams lead to a different final
state. 

Ignoring the contributions from weak boson emission thus is justified if
a well specified exclusive final state is considered. In experiment this
is rarely the case. It is therefore necessary to carefully consider the
role of weak boson emission when calculating the ${\cal O}(\alpha)$ EW
radiative corrections to hadron collider processes. Qualitatively, one
expects the cross section of weak boson emission processes to grow
asymptotically with $\log^2(\hat s/M_{W,Z}^2)$. In a hypothetical
process where one sums/averages over all electroweak charges
in the final/initial state, the $\log^2(\hat s/M_{W,Z}^2)$ terms arising
from the ${\cal O}(\alpha)$ virtual weak corrections and from weak boson
emission diagrams  
cancel at the parton level in the limit $\sqrt{\hat s}\gg M_{W,Z}$. 
In processes of 
practical interest, such as inclusive jet or isolated photon production,
only a partial sum over the electroweak charges is performed, resulting in
a partial cancellation of the ${\cal O}(\alpha)$ virtual weak
corrections and the 
contributions from weak boson emission. Details depend on the process
and the experimental requirements.

In this paper I examine in detail the role of weak boson emission in
those hadron collider processes for which the ${\cal O}(\alpha)$ virtual
weak corrections 
are known. In each case I determine how large the cross sections of the weak
emission processes are compared with the Born cross section as a
function of kinematic variables which are of experimental
interest. Wherever possible, I compare these results with the effect of
the ${\cal O}(\alpha)$ virtual weak corrections, the statistical and the
(expected) systematic uncertainties.

In case of a charged final state with heavy quarks, weak boson emission
can dramatically modify cross sections. This was pointed out in
Ref.~\cite{Ciafaloni:2006qu} for $s$-channel single top quark
production at the LHC, $pp\to t\bar b,\, \bar tb$. While this process
occurs to lowest order (LO) at ${\cal O}(\alpha^2)$, the corresponding
$W$ emission process, $pp\to t\bar bW^-,\, \bar tbW^+$ occurs at ${\cal
O}(\alpha_s^2\alpha)$ and is completely dominated by $\bar tt$ production. I
show that a similar phenomenon occurs in $t$-channel single top quark
production. However, once realistic cuts are imposed, the effect of the
gluon exchange diagrams is found to be small.

All tree level (NLO QCD) cross sections in this paper are computed using
CTEQ6L1 (CTEQ6M)~\cite{Pumplin:2002vw} parton distribution functions
(PDFs). For the CTEQ6L1 PDF's, the 
strong coupling constant is evaluated at leading order with
$\alpha_s(M_Z^2)=0.130$. The factorization and renormalization scales
are set equal to $M_Z$. Since I mostly consider cross section ratios,
results are quite insensitive to the choice of PDFs and the
factorization and renormalization scales. The Standard Model (SM)
parameters used in all tree-level calculations are~\cite{Mangano:2002ea}
\begin{eqnarray}
\label{eq:input1}
G_{\mu} = 1.16639\times 10^{-5} \; {\rm GeV}^{-2}, & \quad & \\
M_Z = 91.188 \; {\rm GeV}, & \quad & M_W=80.419  \; {\rm GeV}, \\
\label{eq:input2} 
\sin^2\theta_W=1-\left({M^2_W\over M_Z^2}\right), & \quad &
\alpha_{G_\mu} = {\sqrt{2}\over\pi}\,G_F \sin^2\theta_W M_W^2,
\end{eqnarray}
where $G_F$ is the Fermi constant, $M_W$ and $M_Z$ are the $W$ and
$Z$ boson masses, $\theta_W$ is the weak mixing angle, and
$\alpha_{G_\mu}$ is the electromagnetic coupling constant in the $G_\mu$
scheme. I use the narrow width approximation for $W$, $Z$ and top quark
decays, unless stated otherwise. Decay correlations are
fully taken into account. Since I use the narrow width approximation,
non-resonant diagrams which yield the same final state as $W$ and $Z$
boson emission with subsequent weak boson decay can be neglected. These
diagrams formally contribute at one order higher in $\alpha$ than the
weak boson emission diagrams.

In Sec.~\ref{sec:sec2} I consider weak boson emission in QCD
processes. Inclusive jet, isolated photon, and $Z+1$~jet production are
examined. Charged and neutral Drell-Yan production are studied in
Sec.~\ref{sec:sec3}. Weak boson emission in di-boson ($W\gamma$,
$Z\gamma$, $WZ$, $ZZ$ and $WW$) production is calculated in
Sec.~\ref{sec:sec4}, and in Sec.~\ref{sec:sec5} this is done for $t\bar
t$ and single top production. I summarize my findings in
Sec.~\ref{sec:sec6}. 

\section{QCD Processes}
\label{sec:sec2}

\subsection{Inclusive jet production}

The measurement of inclusive jet production in hadronic collisions
constitutes a classic test of perturbative QCD. Recent
experimental results from Run~II of the Fermilab Tevatron are described
in Refs.~\cite{jetCDF} and~\cite{jetdzero}. The lowest order process
contributing to inclusive jet production is di-jet production at ${\cal
O}(\alpha_s^2)$. The contributions from tree-level ${\cal
O}(\alpha_s\alpha)$ and ${\cal O}(\alpha^2)$ diagrams~\cite{Baur:1989qt}
and the NLO QCD corrections to di-jet production~\cite{Kunszt:1992tn}
have been known for more than one decade. More recently, the ${\cal
O}(\alpha)$ virtual weak corrections to di-jet production have been
calculated and the tree-level ${\cal
O}(\alpha_s\alpha)$ and ${\cal O}(\alpha^2)$ contributions were included
in the calculation of the inclusive jet cross
section~\cite{Moretti:2005ut}. Photonic corrections are not taken into
account in this analysis.

For their analysis of inclusive jet production, the Tevatron
experiments~\cite{jetCDF,jetdzero} select events which contain one or
more isolated jets in a given transverse momentum ($p_T$) and
pseudo-rapidity ($\eta$) range. There are no constraints on the number
of electrons or muons in the event. However, in order to reduce the
background from cosmic rays, events with large missing transverse
momentum, $p\llap/_T$, are removed. 

The experimental selection criteria imply that ${\cal
O}(\alpha_s\alpha)$ $V+1$~jet ($V=W^\pm,\,Z$) and ${\cal
O}(\alpha_s^2\alpha)$ $V+2$~jet production should be taken into
account when calculating the cross section for inclusive jet
production. Although events with $W\to\ell\nu$ ($\ell=e,\,\mu$) and
$Z\to\bar\nu\nu$ are suppressed due to the $p\llap/_T$ cut, those with
$V\to jj$ fully contribute. 

In order to properly take into account the associated
production of weak bosons with jets up to ${\cal O}(\alpha_s^2\alpha)$,
$V+1$~jet production at NLO QCD has to be calculated. Utilizing the
results of Ref.~\cite{Giele:1993dj}, I have evaluated the contribution
of $V+1$~jet production with $W\to\ell\nu,\, jj$ ($\ell=e,\,\mu,\,\tau$)
and $Z\to\ell^+\ell^-,\,\bar\nu\nu,\, jj$ to the inclusive jet cross
section at NLO in QCD. At least one jet with
\begin{equation}
\label{eq:ptj}
p_T(j) > 25~{\rm GeV~(Tevatron),} \qquad p_T(j) > 50~{\rm GeV~(LHC),}
\end{equation}
and 
\begin{equation}
\label{eq:rapj}
|\eta(j)|<2.5
\end{equation}
is required in the analysis. All jets satisfying Eqs.~(\ref{eq:ptj})
and~(\ref{eq:rapj}) have 
to be isolated from the other particles in the event, except neutrinos, by 
\begin{equation}
\label{eq:jsep}
\Delta R(j,X)>0.4,
\end{equation}
where $\Delta R=[(\Delta\phi)^2+(\Delta\eta)^2]^{1/2}$ is the separation
in pseudo-rapidity-azimuth space, and $X=j,\,\ell$. As cross
checks, LO $V+1$~jet and $V+2$~jet production are calculated. 

Here, and in all other calculations presented in this paper,
$\tau$-lepton decays are not taken into account. Since the
branching ratios for $W\to\tau\nu$ and $Z\to\tau^+\tau^-$ are small,
the approximation of treating $\tau$-leptons on an equal footing with
electrons and muons will change numerical results by at most a few
percent. This is significantly less than the systematic uncertainties
from the choice of PDFs, or the choice of the factorization or
renormalization scale. 

\begin{figure}[th!] 
\begin{center}
\includegraphics[width=13.4cm]{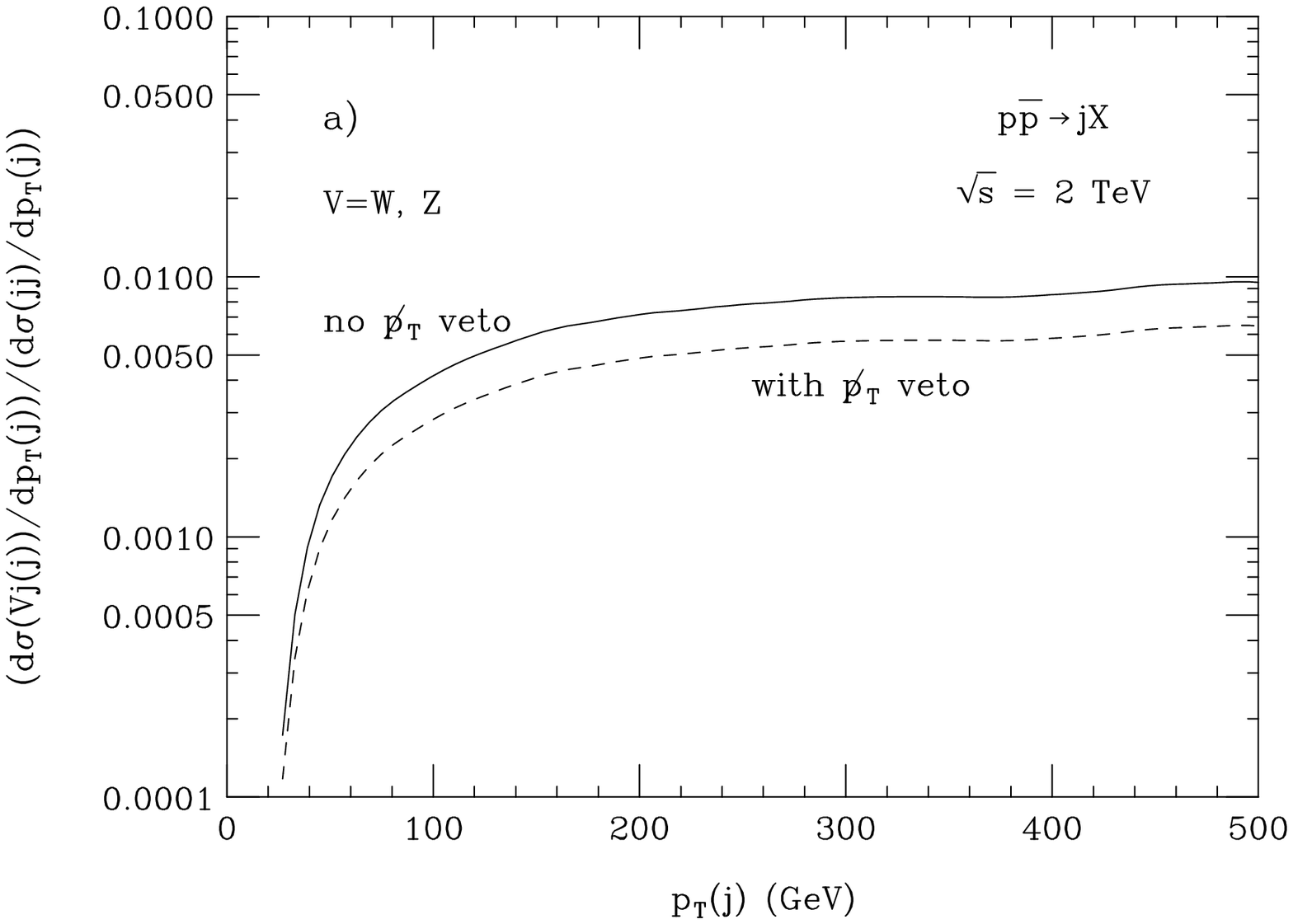} \\[3mm]
\includegraphics[width=13.4cm]{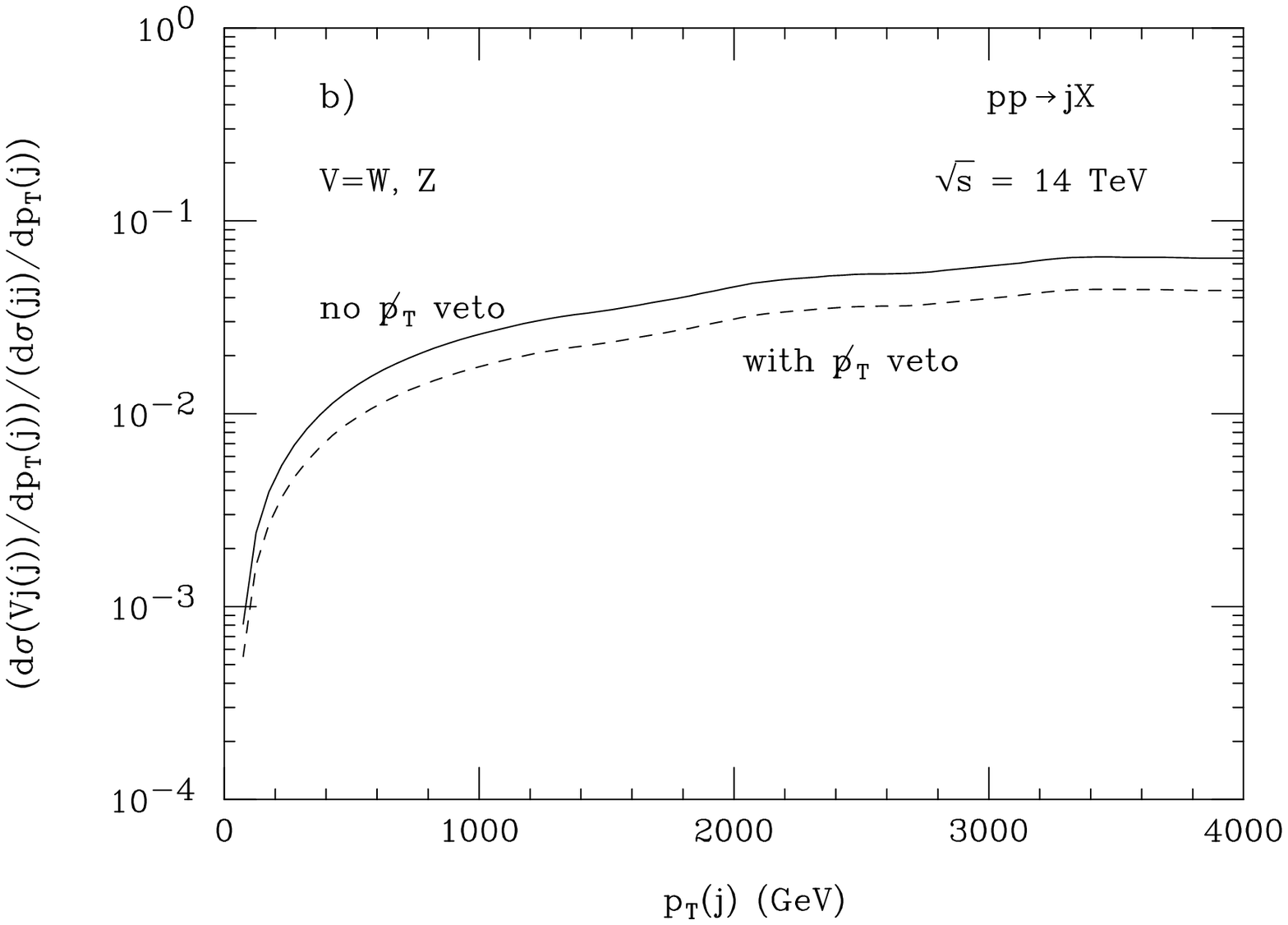}
\vspace*{2mm}
\caption[]{\label{fig:fig1} 
Ratio of the NLO QCD $Vj$ ($V=W^\pm,\, Z$; $W\to\ell\nu,\, jj$,
$Z\to\ell^+\ell^-,\,\bar\nu\nu, \,jj$) and the ${\cal
O}(\alpha_s^2)$ di-jet cross 
section for inclusive jet production as a function of the jet transverse
momentum a) at the Tevatron, and b) at the LHC. 
Results are shown for
two extreme cases: with no $p\llap/_T$ veto imposed (solid line), and
removing all events with non-zero $p\llap/_T$ (dashed line). The cuts
imposed are described in the text.}
\vspace{-7mm}
\end{center}
\end{figure}

In Fig.~\ref{fig:fig1}, I show the ratio of the $Vj(j)$ and the ${\cal
O}(\alpha_s^2)$ di-jet cross section for inclusive jet production as a
function of the jet transverse momentum at the Tevatron and the
LHC. In events with more than one jet satisfying Eqs.~(\ref{eq:ptj})
and~(\ref{eq:rapj}), each jet contributes, {\it ie.} such events produce
multiple entries in the distribution.
Since the $p\llap/_T$ veto imposed depends on experimental details, I
show results for two extreme cases: no $p\llap/_T$ veto (solid line) and
removing all events with non-zero $p\llap/_T$ (dashed line). For a
realistic $p\llap/_T$ veto, the result will fall somewhere in between
the two lines. 

At the Tevatron (LHC), the $Vj(j)$ to LO di-jet cross section ratio
asymptotically approaches $0.6-1.0\%$ ($4.5-6.5\%$) at large values of
$p_T(j)$. $V+1$~jet production contributes significantly only for
$p_T(j)<200-300$~GeV; for large jet transverse momenta $Vjj$ production
dominates. This is to be expected: at large values of $p_T(j)$, soft and
collinear logarithms appear in $Vjj$ production. This is not the case in 
$p\,p\hskip-7pt\hbox{$^{^{(\!-\!)}}$} \to Vj$, where the $W$ or $Z$
boson balances the jet in transverse momentum. At very large jet
transverse momentum, the separation cut of Eq.~(\ref{eq:jsep}) limits
the relative growth of the $Vj(j)$ cross section.

The calculation performed here approximates the QCD
corrections associated with the hadronic decays of the $W$ or $Z$
boson by using a branching fraction for $V\to jj$ which takes into
account NLO QCD corrections. This approximation does not correctly treat
contributions from $Vj$ production with $V\to jjj$. A correct treatment
of the final state QCD corrections is expected to modify the
result presented here at the $10-30\%$ level for $p_T(j)<200$~GeV, but
will have a negligible effect at 
higher jet transverse momenta. Note that, except for $p_T(j)<M_V/2$,
contributions from $p\,p\hskip-7pt\hbox{$^{^{(\!-\!)}}$} \to V\to 4j$
are also negligible (they are of ${\cal O}(\alpha_s^2\alpha^2)$). They are
not included in the calculation presented here.

The contributions of the weak boson emission processes
$p\,p\hskip-7pt\hbox{$^{^{(\!-\!)}}$} \to Vj(j)$ to the inclusive jet
cross section should be compared with
those of the tree level ${\cal O}(\alpha_s\alpha+\alpha^2)$ diagrams and
the ${\cal O}(\alpha)$ virtual weak corrections
calculated in Ref.~\cite{Moretti:2005ut}. Since I am using somewhat
different input parameters, an exact comparison is not
possible. Nevertheless, it is instructive to list the relative size of
the $Vj(j)$ contributions, $\delta(Vj(j))$, and the combined 
tree level ${\cal O}(\alpha_s\alpha+\alpha^2)$ and the ${\cal
O}(\alpha_s^2\alpha)$ virtual one-loop weak corrections,
$\delta$(1-loop), side-by-side. The results are shown in
Table~\ref{tab:tab1}. 
\begin{table}[t!]
\caption{Relative size of the $Vj(j)$ contributions, $\delta(Vj(j))$, and
the combined 
tree level ${\cal O}(\alpha_s\alpha+\alpha^2)$ and the ${\cal
O}(\alpha_s^2\alpha)$ virtual one-loop weak corrections, $\delta$(1-loop), 
to inclusive jet production at the Tevatron and LHC as a function of the
jet transverse momentum. Cross sections are normalized to the ${\cal
O}(\alpha_s^2)$ cross section. The results for the tree level ${\cal
O}(\alpha_s\alpha+\alpha^2)$ and ${\cal O}(\alpha_s^2\alpha)$ one-loop
virtual weak corrections are taken from Ref.~\protect\cite{Moretti:2005ut}.
Results for 
$\delta(Vj(j))$ with a $p\llap/_T$ veto are given in parenthesis.}
\label{tab:tab1}
\vskip 5.mm
\begin{tabular}{c|c|c||c|c|c}
\multicolumn{3}{c||}{Tevatron} & \multicolumn{3}{c}{LHC} \\
\tableline
$p_T(j)$ (GeV) & $\delta$(1-loop) (\%) & $\delta(Vj(j))$ (\%) & 
$p_T(j)$ (GeV) & $\delta$(1-loop) (\%) & $\delta(Vj(j))$ (\%) \\
\tableline
100 & -0.36 & 0.41 (0.28) & 1000 & -9 &  2.5 (1.7) \\
550 & -6.9 & 1.1 (0.73) & 4000 & -24 & 6.5 (4.4)
\end{tabular}
\end{table}

As expected, the contributions from weak boson
emission partially cancel the effects of the one-loop virtual weak
corrections. At the Tevatron, at small jet transverse momenta,
$\delta(Vj(j))$ and $\delta$(1-loop) approximately cancel. At large
values of $p_T(j)$, weak boson emission reduces the 
effect of the ${\cal O}(\alpha)$ virtual weak corrections and the tree
level ${\cal 
O}(\alpha_s\alpha+\alpha^2)$ diagrams by $10-13\%$. At the LHC, weak
boson emission diagrams play a larger role; here
$\delta(Vj(j))/\delta$(1-loop) is in the range $0.2-0.3$.

In order to determine whether the combined effect of the ${\cal
O}(\alpha)$ virtual weak corrections and the contributions from tree
level ${\cal 
O}(\alpha_s\alpha+\alpha^2)$ and weak boson emission diagrams need to be
taken into account for the analysis of Tevatron and LHC inclusive jet
data, the 
results shown in Table~\ref{tab:tab1} have to be compared with the
statistical, systematic and other theoretical uncertainties. At the
Tevatron, the systematic and PDF uncertainties increase from about 10\%
at low $p_T(j)$ to $\sim 40\%$ at
$p_T(j)=500$~GeV~\cite{jetCDF,jetdzero}. Uncertainties from higher order
QCD corrections are $\leq 10\%$ for the $p_T(j)$ range
considered~\cite{jetdzero}. Except for the highest jet
transverse momenta, electroweak corrections thus should be negligible at
the Tevatron. At the LHC one expects systematic and PDF uncertainties of
$10-20\%$ each for $p_T(j)\leq 1$~TeV, and $\sim 50\%$ at
$p_T(j)=4$~TeV~\cite{cmstdr}, the highest jet $p_T$ which can be reached
with an integrated luminosity of 100~fb$^{-1}$. Electroweak radiative
corrections to 
inclusive jet production thus will be relevant for data analysis at the
LHC. 

\subsection{Isolated photon production}

Isolated photon production in hadronic collisions has been another important
tool for probing QCD in the past. It has also presented theoretical
challenges, in particular at fixed target energies (see
Ref.~\cite{Aurenche:2zw} for a recent theoretical review). The most
recent measurements of the Tevatron experiments are described in
Refs.~\cite{Acosta:2002ya} and~\cite{Abazov:2005wc}. The lowest order
process contributing to isolated photon production is
$p\,p\hskip-7pt\hbox{$^{^{(\!-\!)}}$} \to \gamma j$ at ${\cal
O}(\alpha_s\alpha)$. 

For their isolated photon analysis, the Tevatron
experiments select events with a high $p_T$, isolated
photon. Backgrounds from cosmic rays and $W$ decays are reduced by
rejecting events with large $p\llap/_T$. There are no requirements on
the number of charged leptons or jets in the event. In the following, I
assume that similar selection criteria will be used at the LHC. 

The ${\cal O}(\alpha)$ virtual weak corrections to
$p\,p\hskip-7pt\hbox{$^{^{(\!-\!)}}$} \to \gamma j$ were calculated in
Ref.~\cite{Kuhn:2005gv,Maina:2004rb}. Since there are no restrictions on
the number of jets or charged leptons in isolated photon events, 
tree level ${\cal O}(\alpha_s\alpha^2)$ $V\gamma j$ ($V=W^\pm,\,Z$) 
production should be included in the calculation. Likewise, ${\cal
O}(\alpha^2)$ $V\gamma$ production should also be taken into account. 
$V\gamma j$ production is known to exhibit a logarithmic enhancement
factor which is similar to that found in the ${\cal O}(\alpha)$ virtual
weak corrections. In the limit of large 
photon transverse momenta, $p_T(\gamma)\gg M_V$, the $q_1g\to V\gamma
q_{1,2}$ differential cross section can be written in the
form~\cite{Frixione:1992pj,Baur:1993ir} 
\begin{equation}
\label{eq:nlo}
d\hat\sigma(q_1g\to V\gamma q_{1,2})=d\hat\sigma(q_1g\to \gamma q_1)\,
{\alpha\over 4\pi\sin^2\theta_W}\,\log^2\left({p_T^2(\gamma)\over
M_V^2}\right).
\end{equation}

\begin{figure}[th!] 
\begin{center}
\includegraphics[width=13.4cm]{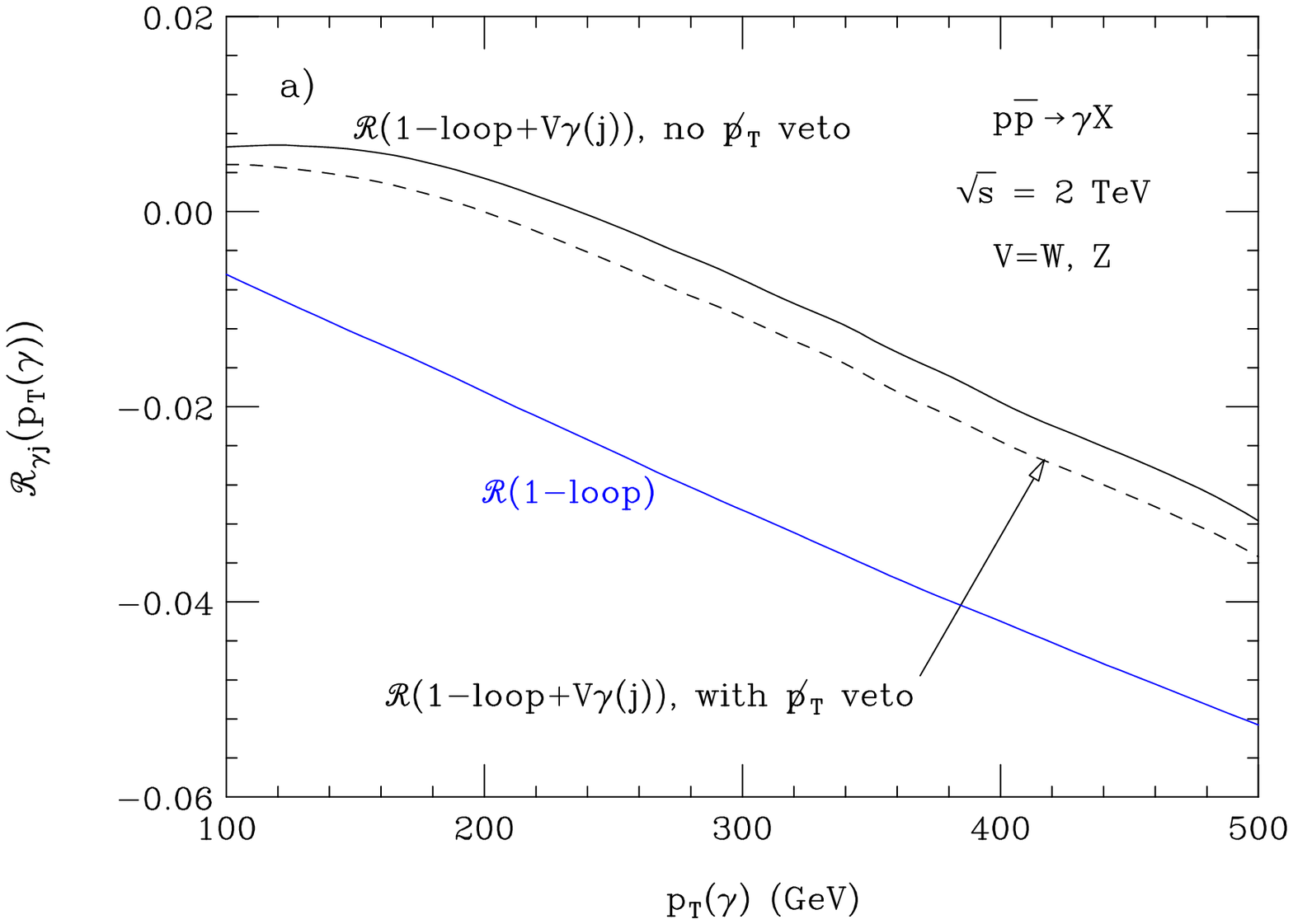} \\[3mm]
\includegraphics[width=13.4cm]{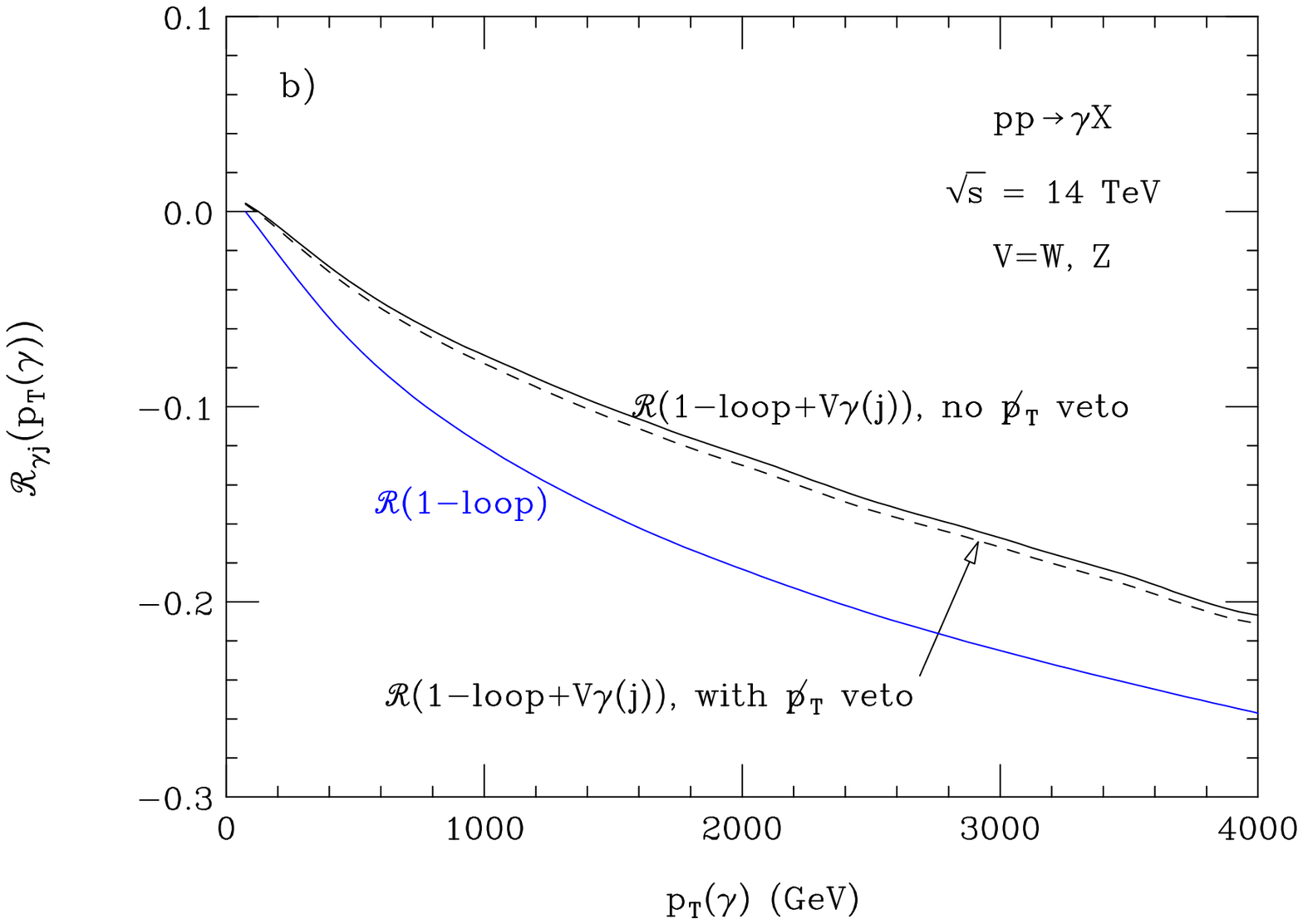}
\vspace*{2mm}
\caption[]{\label{fig:fig2} 
Relative correction with respect to the LO $\gamma j$ cross section,
${\cal R}_{\gamma j}$, as a function of the photon transverse momentum,
$p_T(\gamma)$, for a) the Tevatron and 
b) the LHC. The blue curve shows the result if only the ${\cal
O}(\alpha)$ virtual weak corrections of
Ref.~\protect{\cite{Kuhn:2005gv}} are taken into
account. The black dashed (solid) curve shows ${\cal R}$ if in addition
$V\gamma (j)$ production is included and a (no) $p\llap/_T$ veto is
imposed. The definition of ${\cal R}$ and 
the cuts imposed are described in the text.}
\vspace{-7mm}
\end{center}
\end{figure}

The $V\gamma j$ and $V\gamma$ 
contributions can be taken into account simultaneously by computing the
$p\,p\hskip-7pt\hbox{$^{^{(\!-\!)}}$} \to V\gamma$ cross section
including NLO QCD corrections. Using the results
of~\cite{Baur:1993ir,Ohnemus:1994qp,Baur:1997kz} I have calculated how
NLO QCD $V\gamma$ production modifies the isolated photon cross section
at the Tevatron and LHC. Figure~\ref{fig:fig2}
shows the relative correction ${\cal R}_{\gamma j}(p_T(\gamma))$ with
respect to the LO $\gamma j$ cross section as a function of the photon
transverse momentum. Here, the relative correction is defined as
\begin{equation}
\label{eq:cor}
{\cal R}_Y(X)={d\sigma/dX\over d\sigma^{LO}(Y)/dX}
- 1
\end{equation}
with $X$ being the kinematic variable considered, and $Y$ the final
state of the LO process. 

In Fig.~\ref{fig:fig2}, photons are required to have  
\begin{equation}
|\eta(\gamma)|<2.5
\end{equation}
and be isolated from jets and charged leptons by a distance
\begin{equation}
\Delta R(\gamma,j)>0.4, \qquad \Delta R(\gamma,\ell)>0.4.
\end{equation}
The blue curve shows ${\cal R}_{\gamma j}(p_T(\gamma))$ if only the
${\cal O}(\alpha)$ virtual weak
corrections~\cite{Kuhn:2005gv,Maina:2004rb} are taken into account. It
has been obtained by incorporating the leading ${\cal O}(\alpha)$ virtual weak
corrections of Ref.~\cite{Kuhn:2005gv} into a parton level
$p\,p\hskip-7pt\hbox{$^{^{(\!-\!)}}$} \to \gamma j$ Monte Carlo program,
and by parameterizing the remaining corrections. 

The solid black curve in Fig.~\ref{fig:fig2} displays the result if NLO QCD
$V\gamma$ production with 
$V\to jj$, $W\to\ell\nu$, $Z\to\ell^+\ell^-$ and $Z\to\bar\nu\nu$ is
also included and no
$p\llap/_T$ veto is imposed. The dashed curve finally shows the relative
correction if both ${\cal O}(\alpha)$ virtual weak corrections and NLO QCD
$V\gamma$ production 
are included, however, events are required to have
\begin{equation}
\label{eq:veto}
p\llap/_T\leq 5~{\rm GeV}^{1/2}\sqrt{\sum p_T}.
\end{equation}
Here the sum extends over all particles except neutrinos. The
$p\llap/_T$ veto is seen to have only a relatively small effect. 

Since the $V\gamma$ two body final state does not have any soft or
collinear enhancement factors, it contributes significantly only for
$p_T(\gamma)\leq 200-300$~GeV. This is also the case for $V\gamma$
production with $V\to jjj$, which is not included in the calculation
presented here.

Figure~\ref{fig:fig2}a shows that, at the Tevatron, weak boson emission
effects in isolated photon production essentially
cancel the corrections from the ${\cal O}(\alpha)$ virtual weak diagrams
for photon transverse momenta up to about 200~GeV. At 
$p_T(\gamma)=500$~GeV, they reduce ${\cal R}_{\gamma j}(p_T(\gamma))$
from $-5.2\%$ to $-(3.2-3.5)\%$, {\it ie.} by $30-40\%$. 

At the LHC, with an integrated luminosity of
10~fb$^{-1}$ (100~fb$^{-1}$), it should be possible to measure the
photon transverse momentum distribution for values up to 1.5~TeV
(2.0~TeV). For $p_T(\gamma)=2.0$~TeV, the
combined ${\cal O}(\alpha)$ virtual weak corrections and 
weak boson emission effects reduce the LO $\gamma j$ cross section by
about 13\%, compared with 19\% if the $V\gamma j$ diagrams are
ignored (see Fig.~\ref{fig:fig2}b). The leading ${\cal
O}(\alpha_s\alpha^3)$ two-loop weak corrections~\cite{Kuhn:2005gv} and
${\cal O}(\alpha_s\alpha^2)$ weak boson emission have a very similar
numerical effect on the isolated photon cross section. Thus, when the
leading ${\cal O}(\alpha_s\alpha^3)$ two-loop weak corrections are also
taken into account, weak radiative corrections reduce 
the LO $\gamma j$ cross section only by about 7\% at $p_T(\gamma)=2.0$~TeV.

At the Tevatron, the combined ${\cal O}(\alpha)$ virtual weak
corrections and contributions from weak boson emission do not exceed
3.5\% for photon transverse momenta $p_T(\gamma)\leq 500$~GeV. The
current D\O\ Run~II analysis~\cite{Abazov:2005wc} covers the region
$p_T(\gamma)\leq 300$~GeV. In this region the systematic error varies
between 10 and 20\%, and is always larger than the statistical
uncertainty. The systematic uncertainty decreases with increasing photon
transverse momentum. 
Weak radiative corrections thus will not be important in
isolated photon production at the Tevatron. So far, there are no
estimates of the systematic uncertainties in $pp\to\gamma X$ at the LHC.
However, the results shown in Fig.~\ref{fig:fig2}b show that, unless the
systematic uncertainties are much larger than at the Tevatron, weak
radiative corrections should be taken into account when analyzing
isolated photon production at the LHC.

Weak boson emission effects would be very much reduced if one were to
measure the cross section for exclusive $\gamma+1$~jet production instead of
the inclusive isolated photon cross section. $\gamma+1$~jet
production may be useful for calibrating jet energies at the LHC~\cite{roda}.

\subsection{$Z+1$~jet production}

$Z+1$~jet production is the dominant contribution to $Z$ boson
production at large transverse momentum. The Tevatron
experiments have not yet reported results on the $p_T(Z)$ distribution
for $p_T(Z)>50$~GeV from Run~II. Run~I measurements are described in
Refs.~\cite{Affolder:1999jh} and~\cite{Abbott:1999yd}. $Z$ boson events are
selected by requiring an $e^+e^-$ pair which is consistent in invariant
mass with a $Z$ boson. In a measurement of the transverse momentum
distribution of the $Z$ boson no requirements on the number of jets in
the event are made. However, events with more than two charged leptons
are rejected as di-boson candidates. 

The ${\cal O}(\alpha)$ virtual weak corrections to $Z+1$~jet
production were calculated in Ref.~\cite{Maina:2004rb,Kuhn:2004em}. $Z$
boson decays were not taken into account in this calculation. Since the
number of jets is not fixed in a measurement of the $Z$ boson transverse
momentum, ${\cal O}(\alpha_s\alpha^2)$ $ZVj$
production with $V\to jj$ has to be included when calculating weak
radiative corrections to the $Z$ boson transverse momentum
distribution. $ZZj$ 
events with one $Z$ boson decaying into neutrinos may also contribute,
depending on whether events with a substantial amount of $p\llap/_T$ are
allowed by the experimental selection criteria or not. Similar
to the situation encountered in inclusive jet and isolated photon
production, a more complete treatment includes the contributions from
$ZV$, $V\to jj$ production, and utilizes a calculation which includes the
NLO QCD corrections to these processes. To compute the
contributions of $p\,p\hskip-7pt\hbox{$^{^{(\!-\!)}}$} \to ZV$ at NLO
QCD, I have used the results of Refs.~\cite{Baur:1994aj} in the $WZ$
case, and those of Ref.~\cite{Campbell:1999ah}, as implemented in {\tt
MCFM-5.1}~\cite{mcfm}, for $ZZ$ production. These calculations assume
that both weak bosons decay leptonically. To estimate NLO QCD $ZV$,
$V\to jj$, production, I rescale the cross section obtained for leptonic
$V$ decays to 
correct for the higher $V\to jj$ branching ratio. This approximation
does not correctly treat final state NLO QCD corrections, in particular
$ZV$ production with  
$V\to jjj$. However, similar to the situation encountered in inclusive
jet and isolated photon production, final state QCD corrections to $ZV$
production are expected to have a non-negligible effect only at small
values of $p_T(Z)$. 

\begin{figure}[th!] 
\begin{center}
\includegraphics[width=13.4cm]{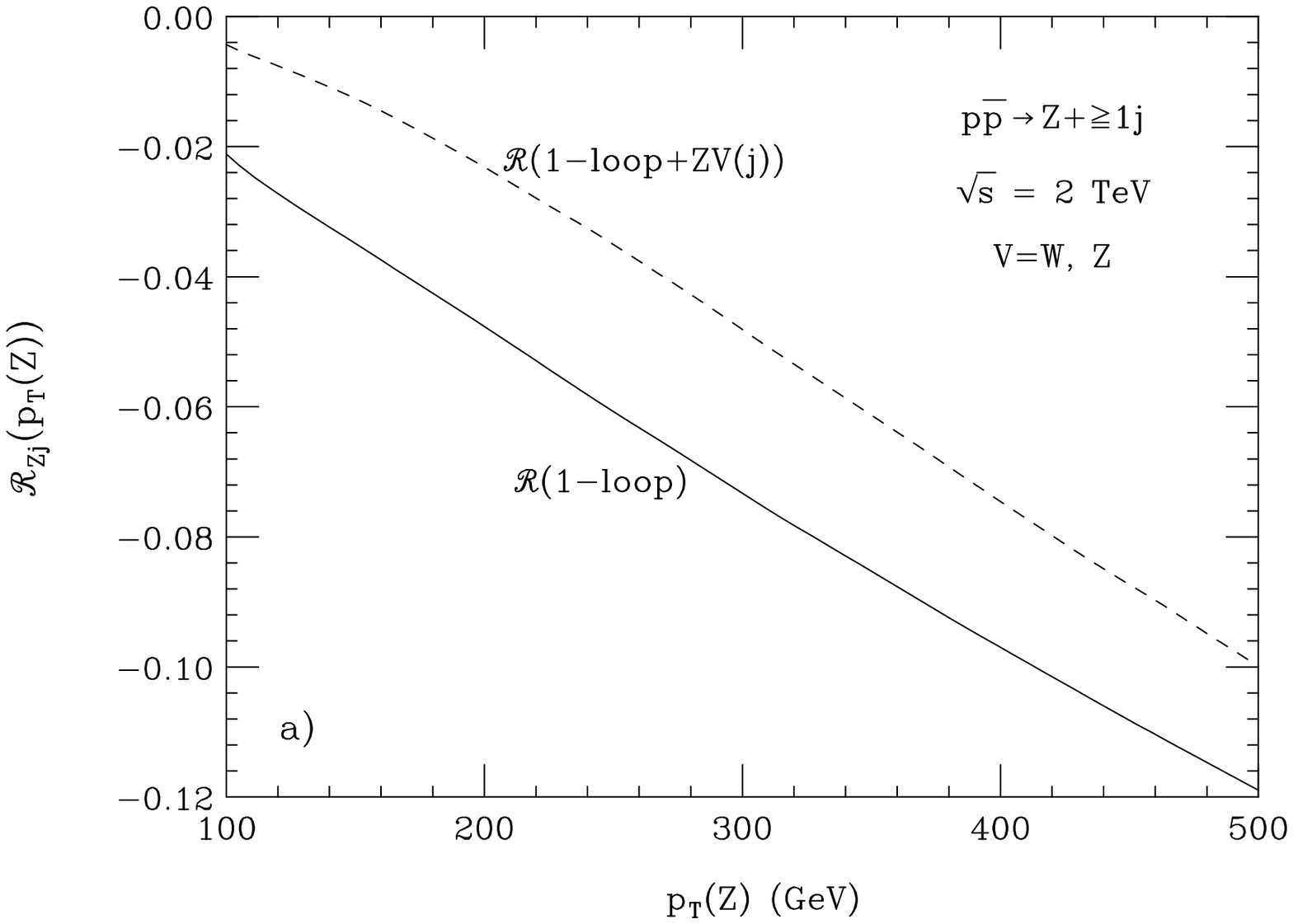} \\[3mm]
\includegraphics[width=13.4cm]{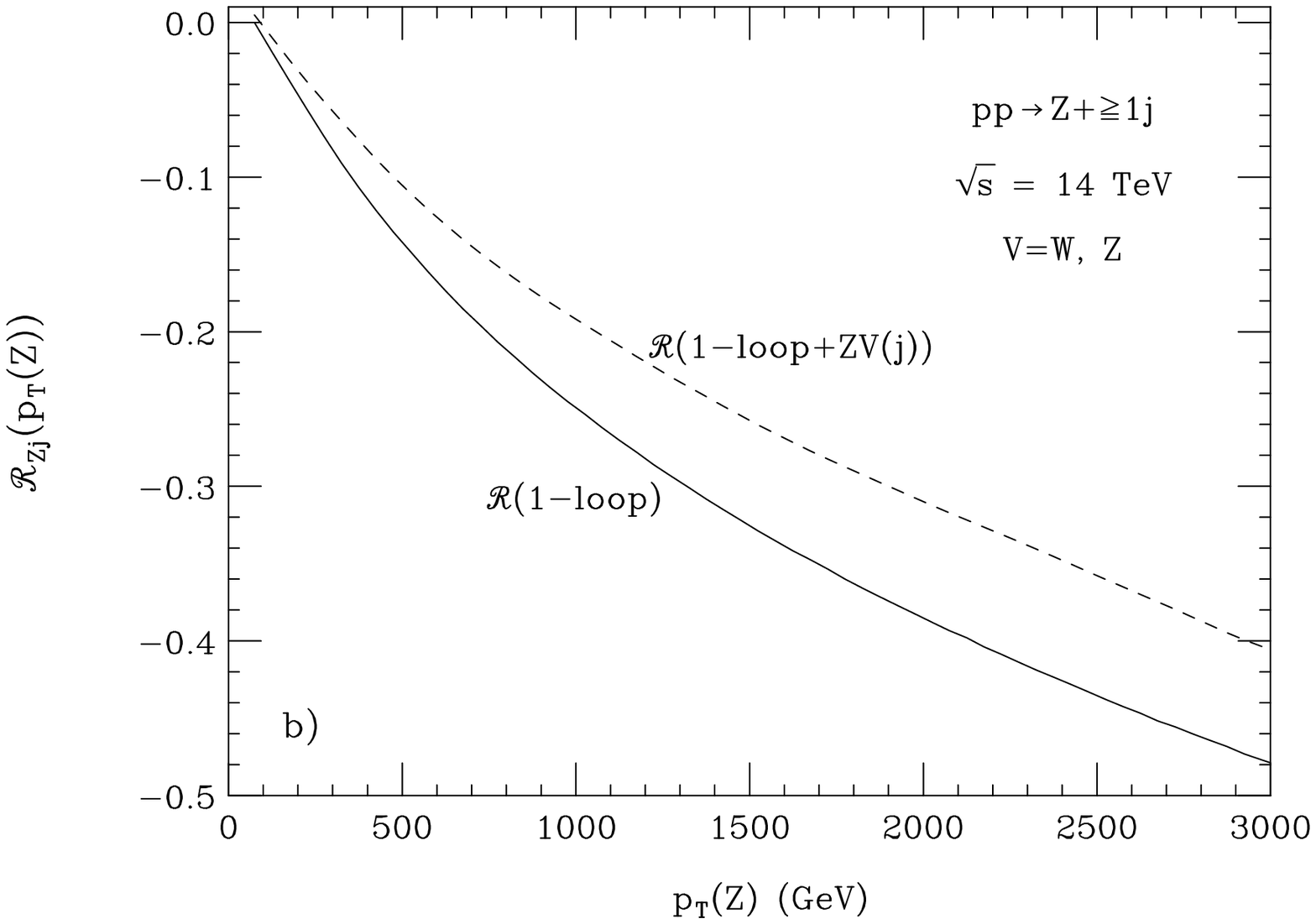}
\vspace*{2mm}
\caption[]{\label{fig:fig3} 
Relative correction with respect to the LO $Z+1$~jet cross section,
${\cal R}_{Zj}$, as a 
function of the $Z$ boson transverse momentum, $p_T(Z)$, for a) the
Tevatron and 
b) the LHC. The solid curve shows the result if only the ${\cal
O}(\alpha)$ virtual weak corrections of
Ref.~\protect{\cite{Kuhn:2004em}} are taken into account. The 
dashed curve shows ${\cal R}_{Zj}(p_T(Z))$ if 
$ZV (j)$ production with $V\to jj$ is included as well. The definition
of ${\cal R}_{Zj}(p_T(Z))$ and the cuts imposed are described in the text.}
\vspace{-7mm}
\end{center}
\end{figure}

The relative correction, ${\cal R}_{Zj}(p_T(Z))$ (see Eq.~(\ref{eq:cor})),
to the lowest order $Z+1$~jet cross section is shown in Fig.~\ref{fig:fig3}.
Here I require that events contain at least one jet with
\begin{equation}
p_T(j) > 25~{\rm GeV~(Tevatron),} \qquad p_T(j) > 50~{\rm GeV~(LHC),}
\end{equation}
and 
\begin{equation}
|\eta(j)|<2.5.
\end{equation}
In addition I impose the $p\llap/_T$ veto of Eq.~(\ref{eq:veto}).

The solid curve shows the result taking only the ${\cal
O}(\alpha)$ virtual weak corrections~\cite{Maina:2004rb,Kuhn:2004em}  
into account. It has been obtained by incorporating the leading ${\cal
O}(\alpha)$ virtual weak 
corrections of Ref.~\cite{Kuhn:2004em} into a parton level
$p\,p\hskip-7pt\hbox{$^{^{(\!-\!)}}$} \to Z j$ Monte Carlo program,
and by parameterizing the remaining corrections. 
The dashed line displays ${\cal R}_{Zj}(p_T(Z))$ if
$ZV(j)$ production with $V\to jj$ is also included in the
calculation. The two-body process
$p\,p\hskip-7pt\hbox{$^{^{(\!-\!)}}$} \to ZV$ contributes significantly
only for small $Z$ boson transverse momenta. 

The ${\cal O}(\alpha)$ virtual weak corrections to $Z+1$~jet production
are considerably 
larger than those found for isolated photon production. At the Tevatron,
weak boson emission increases ${\cal R}_{Zj}(p_T(Z))$ by about 2\% over
the $p_T$ range studied here. In Run~II, CDF and D\O\ should be able to map
out the $p_T(Z)$ distribution up to transverse momenta of
$350-400$~GeV. In this range, the full ${\cal O}(\alpha)$ weak
corrections reduce the LO $Z+1$~jet cross section by $6-8\%$. The weak
radiative corrections are thus of the same size as the expected
systematic uncertainties~\cite{Affolder:1999jh,Abbott:1999yd} which
should dominate over the statistical errors except for the very highest
$p_T(Z)$ bin. 

At the LHC, with $Z\to e^+e^-$, transverse momenta up to 1.0~TeV
(1.5~TeV) can be reached with an integrated luminosity of 10~fb$^{-1}$
(100~fb$^{-1}$). For $p_T(Z)=1.5$~TeV, the ${\cal O}(\alpha)$ virtual
weak corrections 
reduce the LO $Z+1$~jet cross section by about 33\%. Including weak
boson emission decreases the magnitude of ${\cal R}_{Zj}$ 
to 27\%. $W$ and $Z$ boson radiation 
and the leading two-loop weak corrections~\cite{Kuhn:2004em}
have a very similar effect on the $Z+1$~jet cross section at LHC
energies.  
The systematic uncertainties at the LHC and the Tevatron are expected to
be similar~\cite{cmstdr}. It will thus be
important to take into account the full ${\cal O}(\alpha)$
weak corrections, including weak boson emission diagrams, at both the
Tevatron and the LHC.  

\section{Drell-Yan Production}
\label{sec:sec3}

Charged and neutral Drell-Yan production,
$p\,p\hskip-7pt\hbox{$^{^{(\!-\!)}}$} \to \ell\nu$ and
$p\,p\hskip-7pt\hbox{$^{^{(\!-\!)}}$} \to 
\ell^+\ell^-$, at masses and transverse momenta larger the $W$ or $Z$
mass, are tools to search for new heavy gauge bosons~\cite{new}, $W'$ and
$Z'$, and other resonances, such as gravitons in Randall-Sundrum
models~\cite{Randall:1999ee}. The most recent Tevatron Run~II results
for $W'$, $Z'$, and graviton searches are described in
Refs.~\cite{wprime,zprime,grav}. In the charged channel, events are
selected by requiring one charged lepton and large missing transverse
momentum. In the neutral channel, two oppositely charged leptons are
required, and no significant amount of $p\llap/_T$ is allowed. The
number of jets in the event is not restricted in both the charged and the
neutral channel. Weak boson emission, {\it i.e.} $\ell\nu V$ and
$\ell^+\ell^-V$ production with $V\to jj$, may thus contribute
to Drell-Yan production at ${\cal O}(\alpha^3)$. In the charged channel,
$p\,p\hskip-7pt\hbox{$^{^{(\!-\!)}}$} 
\to \ell\nu Z$ with $Z\to\bar\nu\nu$ may also play a role. 

The ${\cal O}(\alpha)$ EW radiative corrections to
$p\,p\hskip-7pt\hbox{$^{^{(\!-\!)}}$} \to \ell\nu$ and
$p\,p\hskip-7pt\hbox{$^{^{(\!-\!)}}$} \to \ell^+\ell^-$ were calculated
in~\cite{Baur:2001ze,kramer,Baur:2004ig,Arbuzov:2005dd,Zykunov:2005tc}.
In the following, I shall use the calculations of
Refs.~\cite{Baur:2001ze} and~\cite{Baur:2004ig}. In addition to the weak
one-loop corrections, these calculations also take into account photonic
corrections. 

The granularity of detectors and the size of electromagnetic showers in
the calorimeter make it difficult to discriminate between electrons and
photons with a small opening angle. In such cases, the four-momentum
vectors of the electron and photon are recombined to an effective
electron four-momentum vector. The exact recombination procedure is
detector dependent. Recombining the electron and
photon four-momentum vectors eliminates the mass singular logarithmic
terms originating from final state photon radiation and strongly reduces
the size of the photonic final state corrections~\cite{Baur:1997wa}.

Muons are identified by hits in the muon chambers and the requirement
that the associated track is consistent with a minimum ionizing
particle. This limits the photon energy for small muon-photon opening
angles. The cut on the photon energy increases the size of the photonic
corrections. The photonic corrections are not of interest for the
following discussion. I therefore focus on final states containing
electrons and impose realistic electron identification
requirements. This minimizes the effect of the photonic
corrections. For $M_T>150$~GeV and $p_T(e)>80$~GeV, the one-loop weak
correction dominate over the photonic corrections.

In the calculations presented in this Section, electrons are required to
have 
\begin{equation}
p_T(e)>25~{\rm GeV}\qquad {\rm and} \qquad |\eta(e)|<2.5.
\end{equation}
The electron identification requirements are taken from
Ref.~\cite{Baur:2004ig}. Electrons also have to be isolated from the
hadronic decay products in $e\nu V$ and $e^+e^- V$ events with $V\to jj$:
\begin{equation}
\Delta R(e,j)>0.4.
\end{equation}
In the charged channel, events also must have
\begin{equation}
p\llap/_T>25~{\rm GeV},
\end{equation}
whereas the $p\llap/_T$ veto of Eq.~(\ref{eq:veto}) is imposed in the
neutral channel. The cross sections for
$p\,p\hskip-7pt\hbox{$^{^{(\!-\!)}}$} \to e\nu V, \, e^+e^-V$ with $V\to
jj$ are calculated using {\tt MadEvent}~\cite{Maltoni:2002qb}. For $e\nu
Z$ and $e^+e^- Z$ production with $Z\to\bar\nu\nu$, the calculation is
based on the complete set of tree level Feynman diagrams contributing to
the $e\nu_e\bar\nu_\ell\nu_\ell$ final state ($\ell=e,\,\mu,\,\tau$).

\begin{figure}[th!] 
\begin{center}
\includegraphics[width=13.4cm]{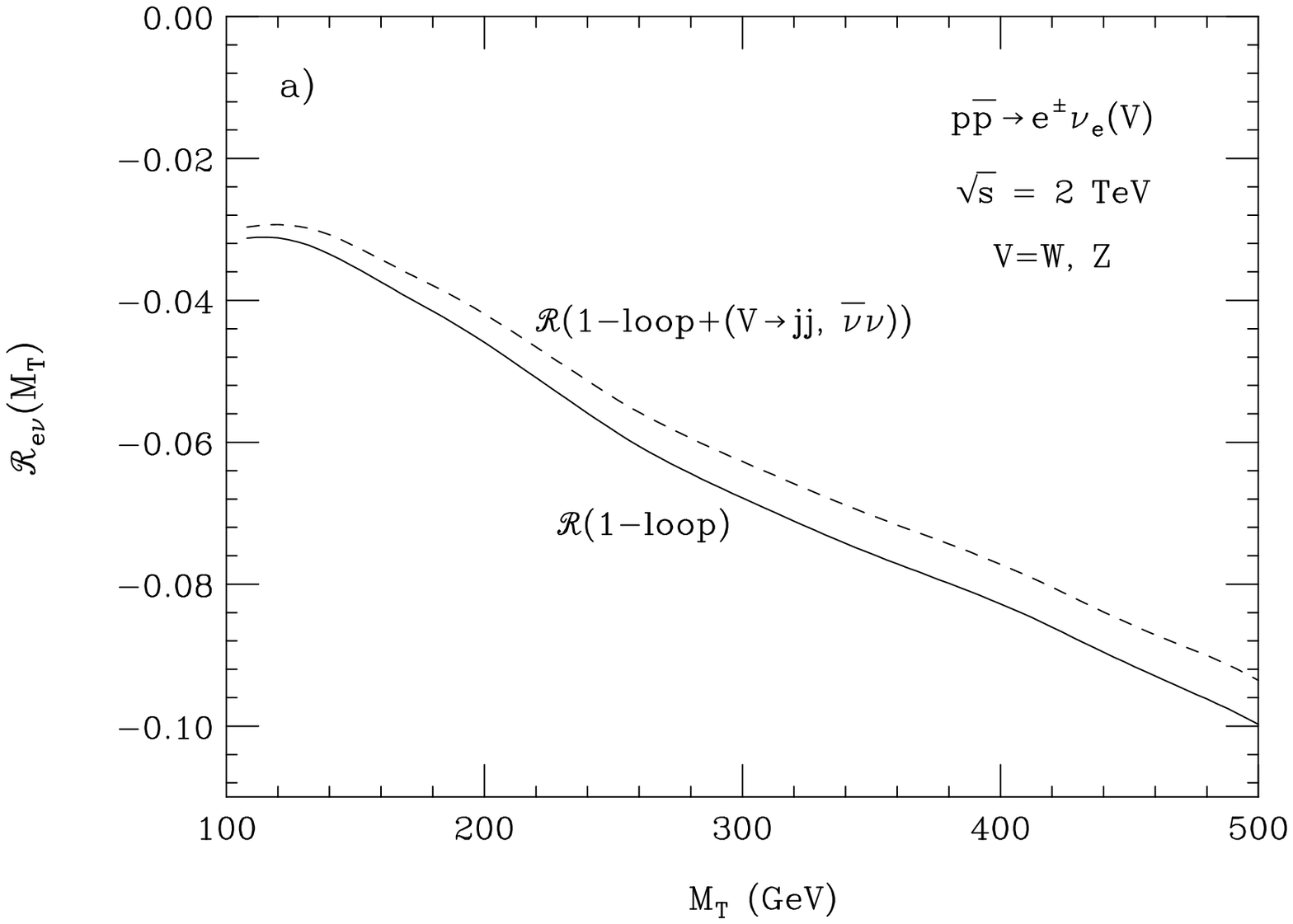} \\[3mm]
\includegraphics[width=13.4cm]{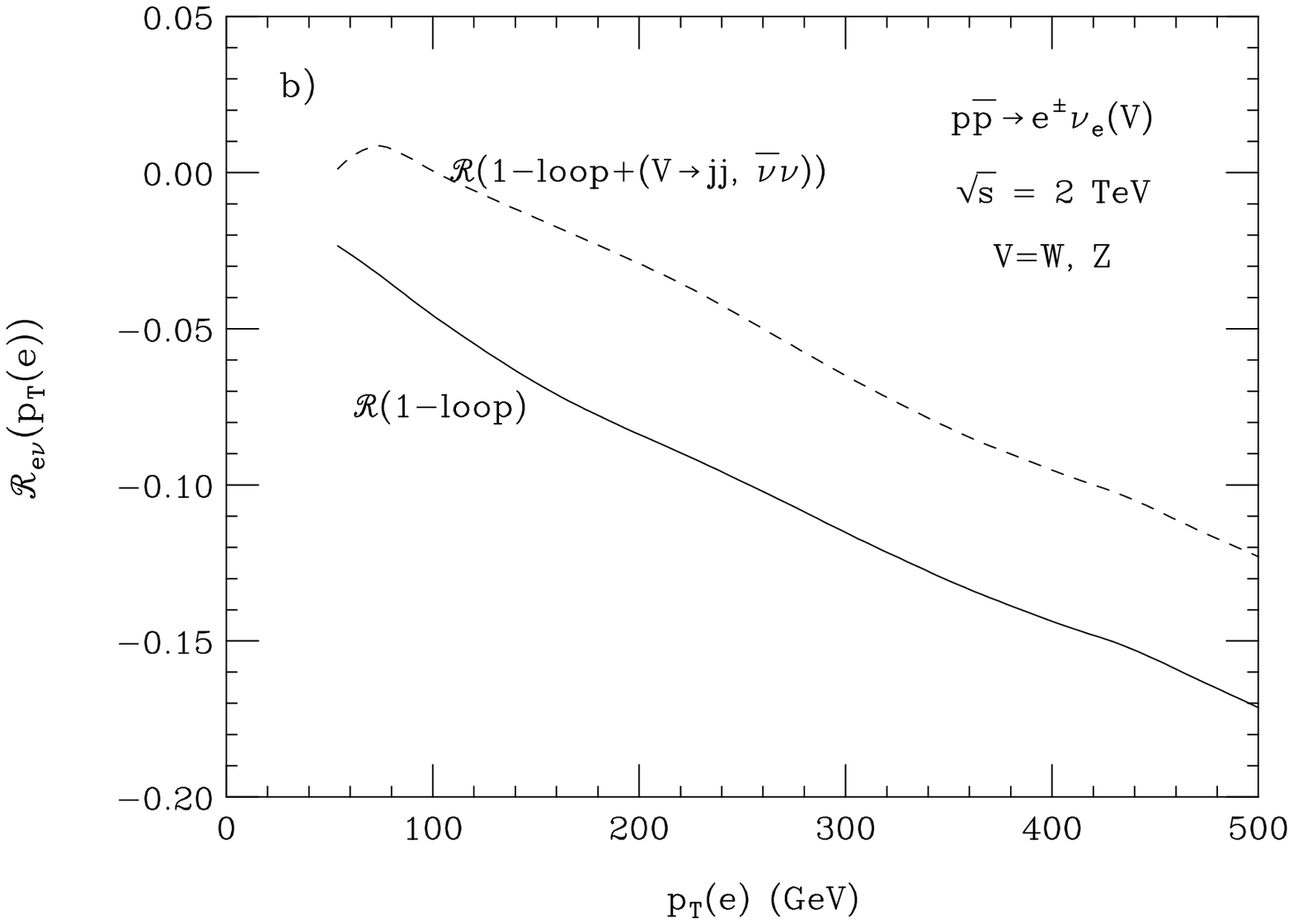}
\vspace*{2mm}
\caption[]{\label{fig:fig4} 
The relative correction with respect to the LO $e\nu$ cross section at the
Tevatron as a 
function a) of the $e\nu$ transverse mass and b) the electron $p_T$. 
The solid curve shows the result if only the ${\cal O}(\alpha)$
corrections of Ref.~\protect{\cite{Baur:2004ig}} are taken into
account. The 
dashed curve shows ${\cal R}_{e\nu}$ if ${\cal O}(\alpha^3)$
$e\nu V$ production with $V\to jj$ and $Z\to\bar\nu\nu$ is included as
well. The definition 
of ${\cal R}_{e\nu}$ and the cuts imposed are described in the text.}
\vspace{-7mm}
\end{center}
\end{figure}

The relative correction to the cross section in the charged channel as a
function of the $e\nu$ 
transverse mass, $M_T$, and the electron transverse momentum, $p_T(e)$,
is shown in Fig.~\ref{fig:fig4}.
The transverse mass is defined by
\begin{equation}
M_T=\sqrt{2p_T(e)p\llap/_T(1-\cos\phi^{ep\!\!\!/_T})},
\end{equation}
where $\phi^{ep\!\!\!/_T}$ is the angle between the electron and the
missing transverse momentum vector in the transverse plane. The solid
line in Fig.~\ref{fig:fig4} shows the result for ${\cal R}_{e\nu}$ for
the one-loop weak and 
photonic ${\cal O}(\alpha)$ corrections of Ref.~\cite{Baur:2004ig}. In
the dashed line, the $e\nu V$ contributions with $V\to jj$ and
$Z\to\bar\nu\nu$ are also included.  

Figure~\ref{fig:fig4}a shows that weak boson emission effects in the
$M_T$ distribution are quite small at the Tevatron. They increase ${\cal
R}_{e\nu}$ by less than 0.01 for the $M_T$ range considered here. The
effect of the weak boson emission diagrams is much more pronounced in
the electron transverse momentum distribution. For $p_T(e)>100$~GeV, $W$
and $Z$ radiation increases ${\cal R}_{e\nu}$ uniformly by about
0.05. The ${\cal O}(\alpha)$ weak one-loop corrections and the
contribution from weak boson emission thus cancel to a significant
degree in the $p_T(e)$ distribution. It is easy to understand why $W$
and $Z$ radiation has a larger effect in the electron transverse
momentum distribution. Since $M_T\leq m(e\nu)$, where $m(e\nu)$ is the
$e\nu$ invariant mass, $m(e\nu)$ is always well above the $W$ resonance
region for the $M_T$ range studied here. On the other hand, for large
$p_T(e)$, the transverse momenta of the electron and the $W$ or $Z$
boson radiated in the event can balance, and the neutrino can be
relatively soft. In this kinematic configuration, the $e\nu$ system
can form an on-shell $W$. In a nutshell, in the $p_T(e)$ distribution,
on-shell $WV$ production contributes, while it does not in the
transverse mass distribution (for $M_T>100$~GeV). 

\begin{figure}[th!] 
\begin{center}
\includegraphics[width=13.4cm]{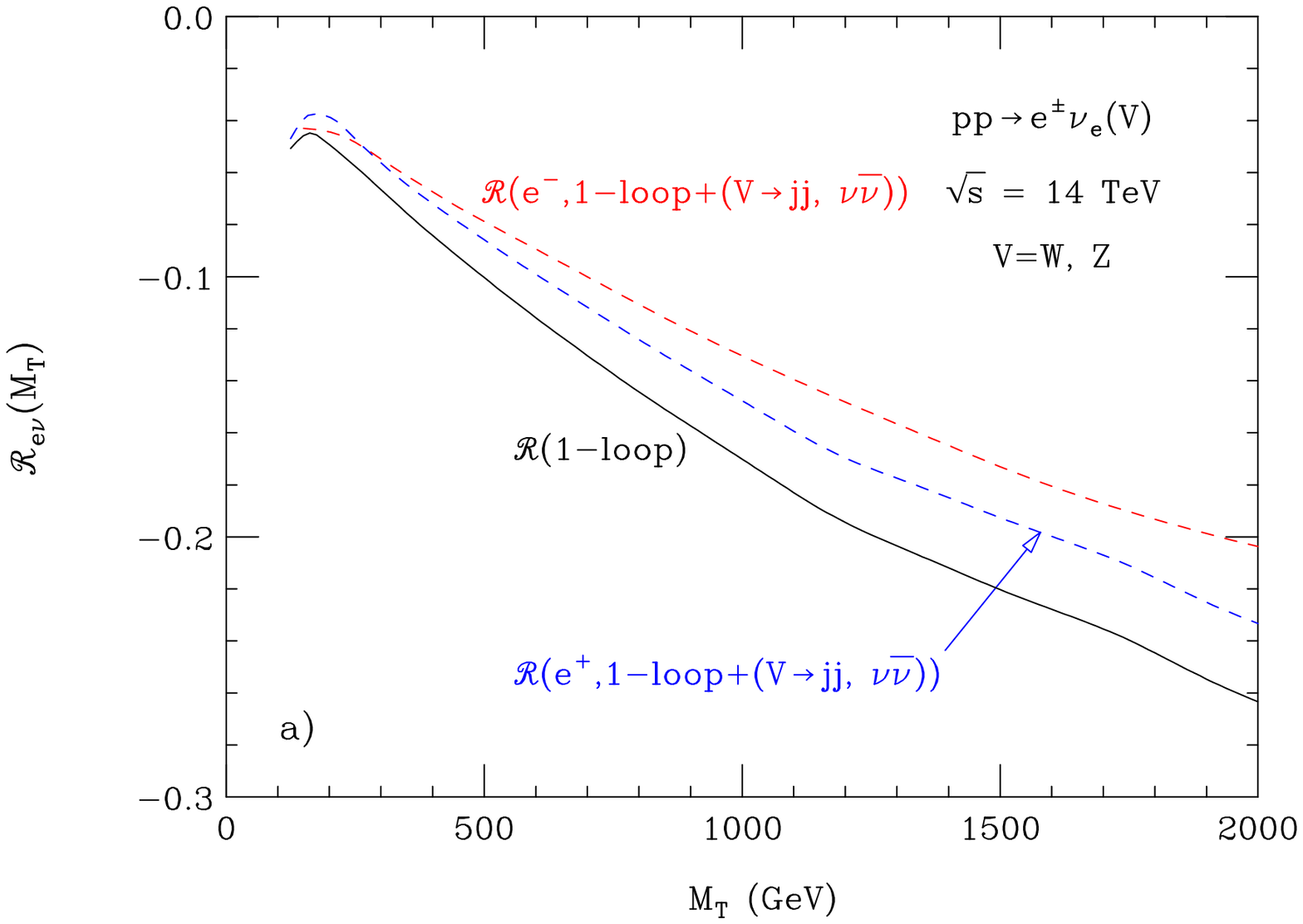} \\[3mm]
\includegraphics[width=13.4cm]{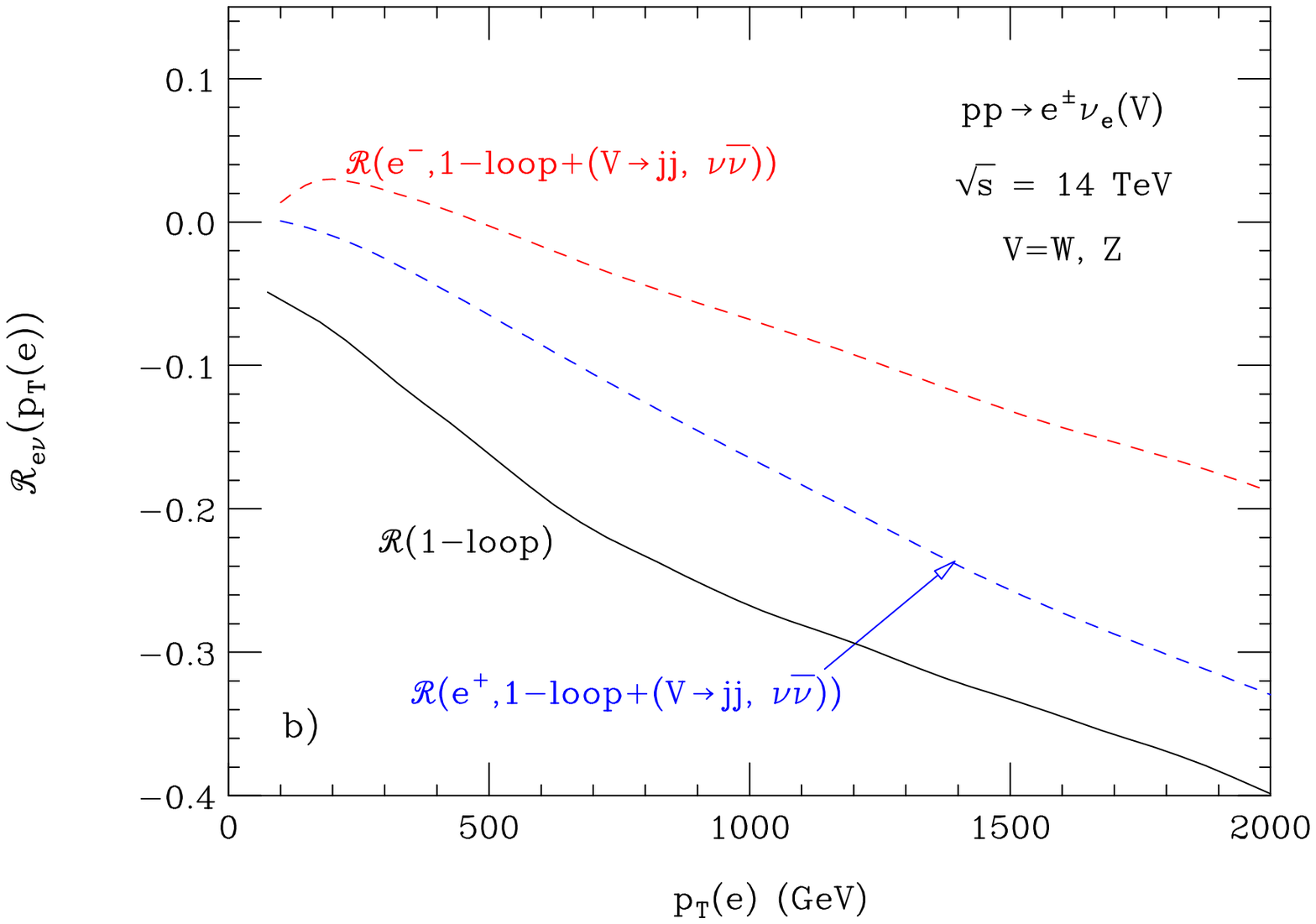}
\vspace*{2mm}
\caption[]{\label{fig:fig5} 
The relative correction with respect to the LO $e\nu$ cross section at the
LHC as a 
function a) of the $e\nu$ transverse mass and b) the electron $p_T$. 
The solid curve shows the result if only the ${\cal O}(\alpha)$
corrections of Ref.~\protect{\cite{Baur:2004ig}} are taken into
account. The 
dashed blue (red) curve shows ${\cal R}_{e\nu}$ in the $e^+\nu$
($e^-\nu$) channel if ${\cal O}(\alpha^3)$
$e\nu V$ production with $V\to jj$ and $Z\to\bar\nu\nu$ is included in
addition. The definition 
of ${\cal R}_{e\nu}$ and the cuts imposed are described in the text.}
\vspace{-7mm}
\end{center}
\end{figure}

${\cal R}_{e\nu}$ as a function of $M_T$ and $p_T(e)$ at the LHC is
shown in Fig.~\ref{fig:fig5}.
As at the Tevatron, weak boson emission effects are more pronounced in
the electron transverse momentum distribution. At the LHC, the cross
sections for $e^+\nu$ and $e^-\bar\nu$ production are different. At
large values of $M_T$ and $p_T(e)$, the $e^+\nu$ cross section is almost
one order magnitude larger. Since the weak one-loop corrections are
proportional to the LO $e^\pm\nu$ cross section and the photonic
corrections are dominated by final state radiation effects, the relative
corrections to the $e^+\nu$ and $e^-\bar\nu$ cross sections due to these
effects are almost equal. They are represented by the black solid lines
in Fig.~\ref{fig:fig5}. Weak boson emission effects are dominated by 
$e^\pm\nu W^\mp$ production which yield equal {\sl cross sections} in
the two cases. Since the LO $e^-\bar\nu$ cross section is much smaller than
the LO $e^+\nu$ rate, $W$ radiation affects ${\cal R}_{e\nu}$ much more
strongly in the $e^-\bar\nu$ channel. 

At $p_T(e)=1.0$~TeV, weak boson
emission reduces ${\cal R}_{e\nu}$ from 28\% to 7\% (17\%) in magnitude
for $e^-\bar\nu$ ($e^+\nu$) production. For an integrated luminosity of
100~fb$^{-1}$, one expects to measure the $p_T(e)$ ($M_T$) distribution
for values up to 1~TeV (2~TeV).

\begin{figure}[th!] 
\begin{center}
\includegraphics[width=13.4cm]{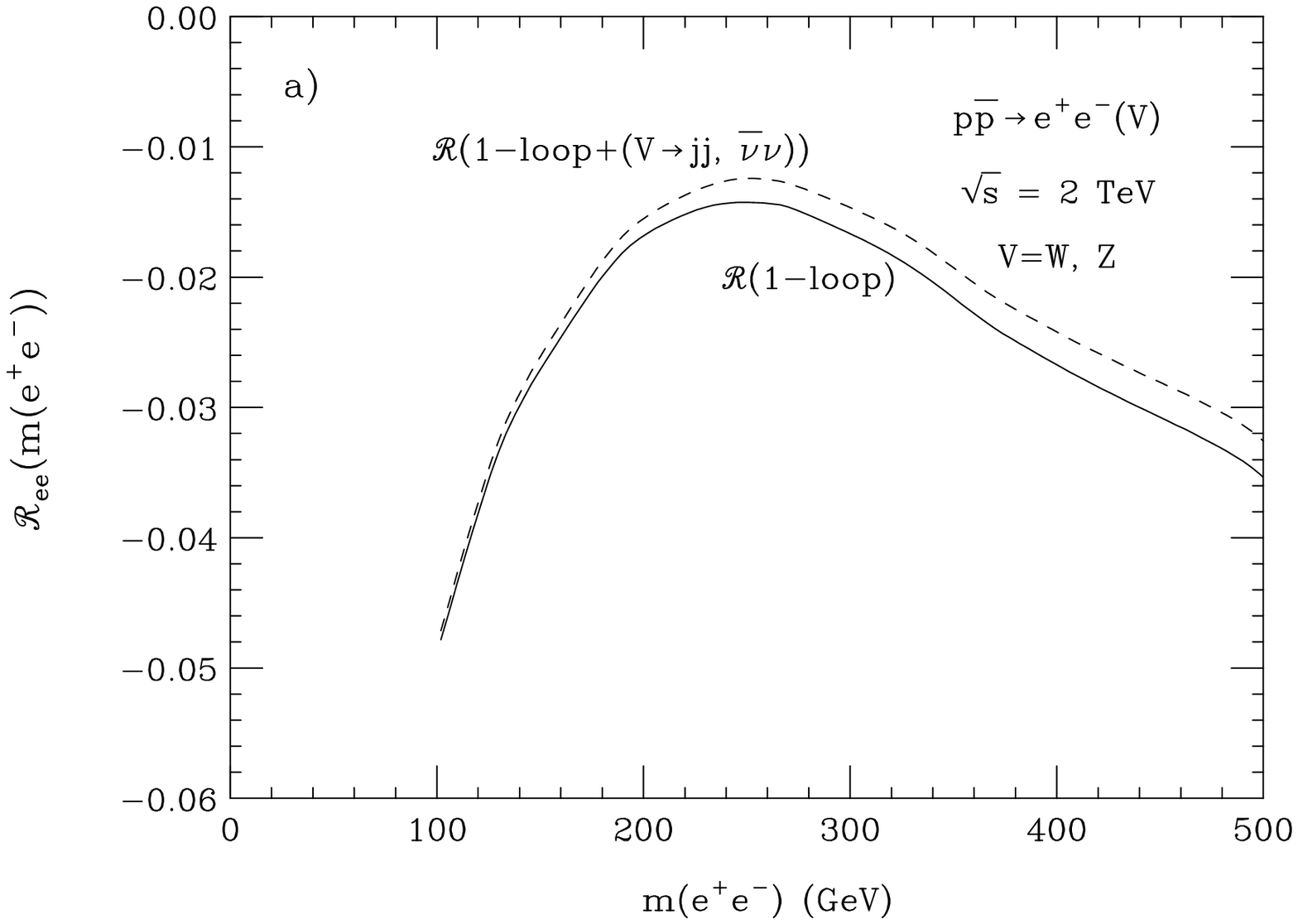} \\[3mm]
\includegraphics[width=13.4cm]{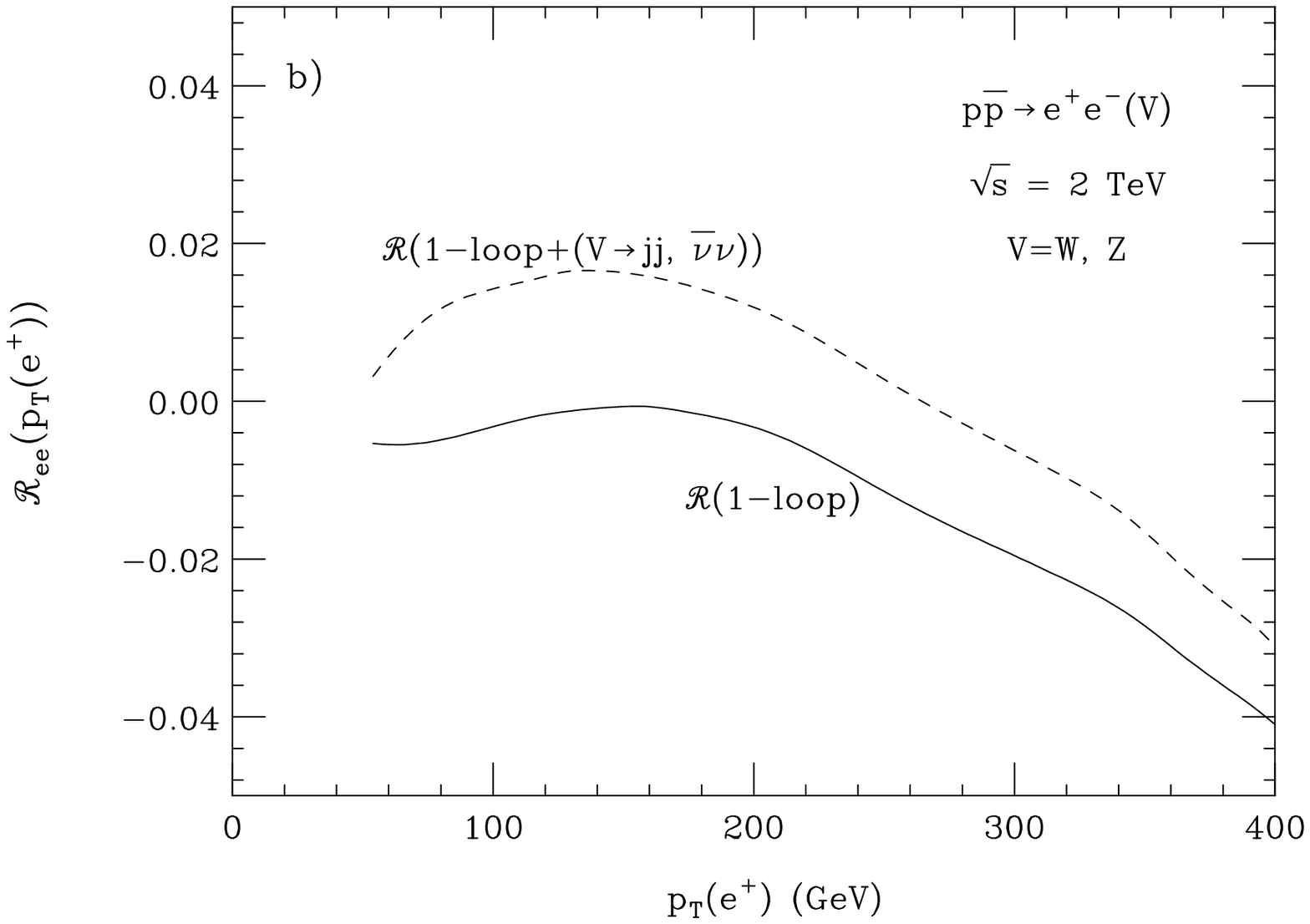}
\vspace*{2mm}
\caption[]{\label{fig:fig6} 
The relative correction with respect to the LO $e^+e^-$ cross section at the
Tevatron as a 
function a) of the $e^+e^-$ invariant mass and b) the positron $p_T$. 
The solid curve shows the result if only the ${\cal O}(\alpha)$
corrections of Ref.~\protect{\cite{Baur:2001ze}} are taken into
account. The 
dashed curve shows ${\cal R}_{ee}$ if ${\cal O}(\alpha^3)$
$e^+e^- V$ production with $V\to jj$ and $Z\to\bar\nu\nu$ is included as
well. The definition 
of ${\cal R}_{ee}$ and the cuts imposed are described in the text.}
\vspace{-7mm}
\end{center}
\end{figure}

\begin{figure}[th!] 
\begin{center}
\includegraphics[width=13.4cm]{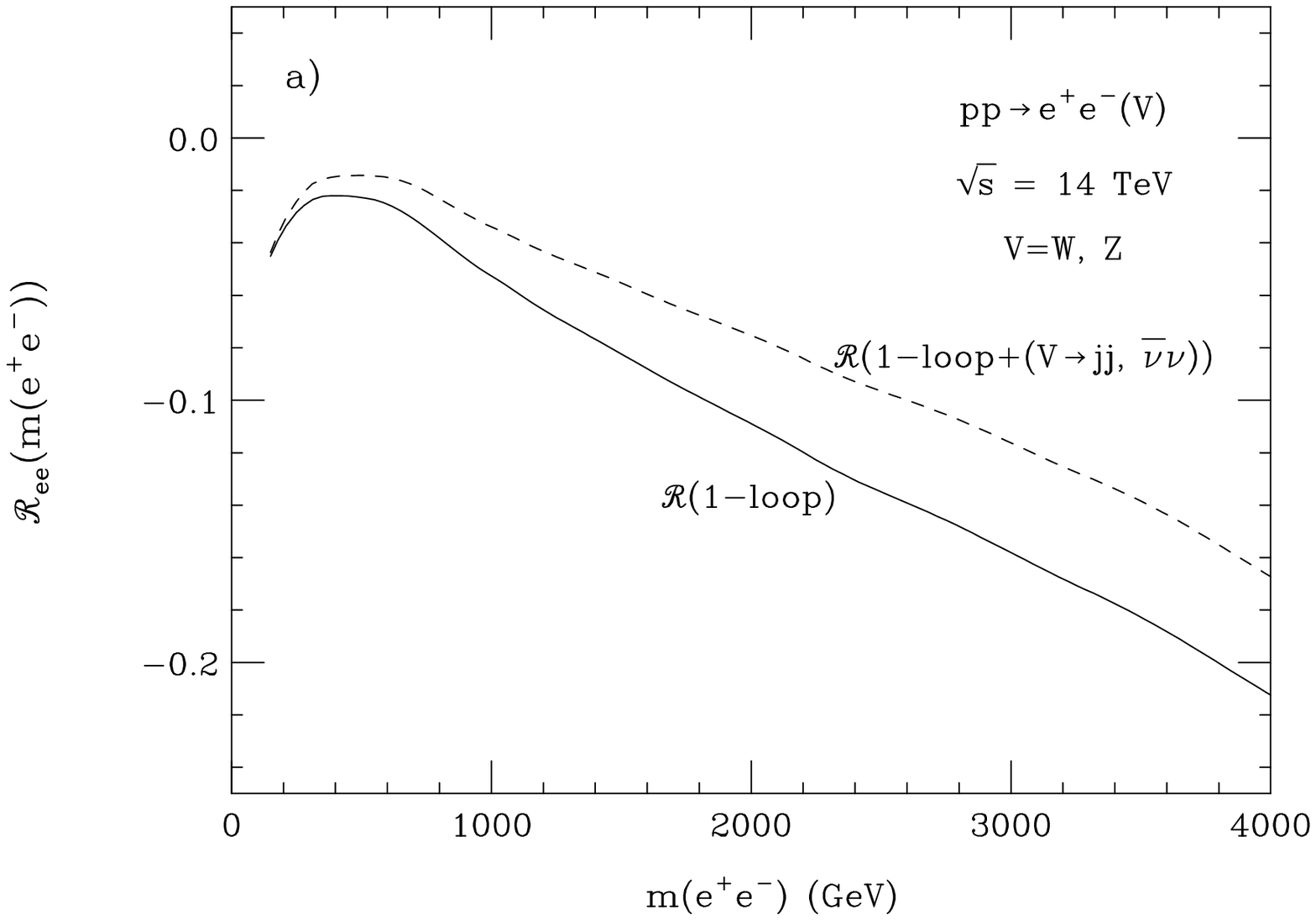} \\[3mm]
\includegraphics[width=13.4cm]{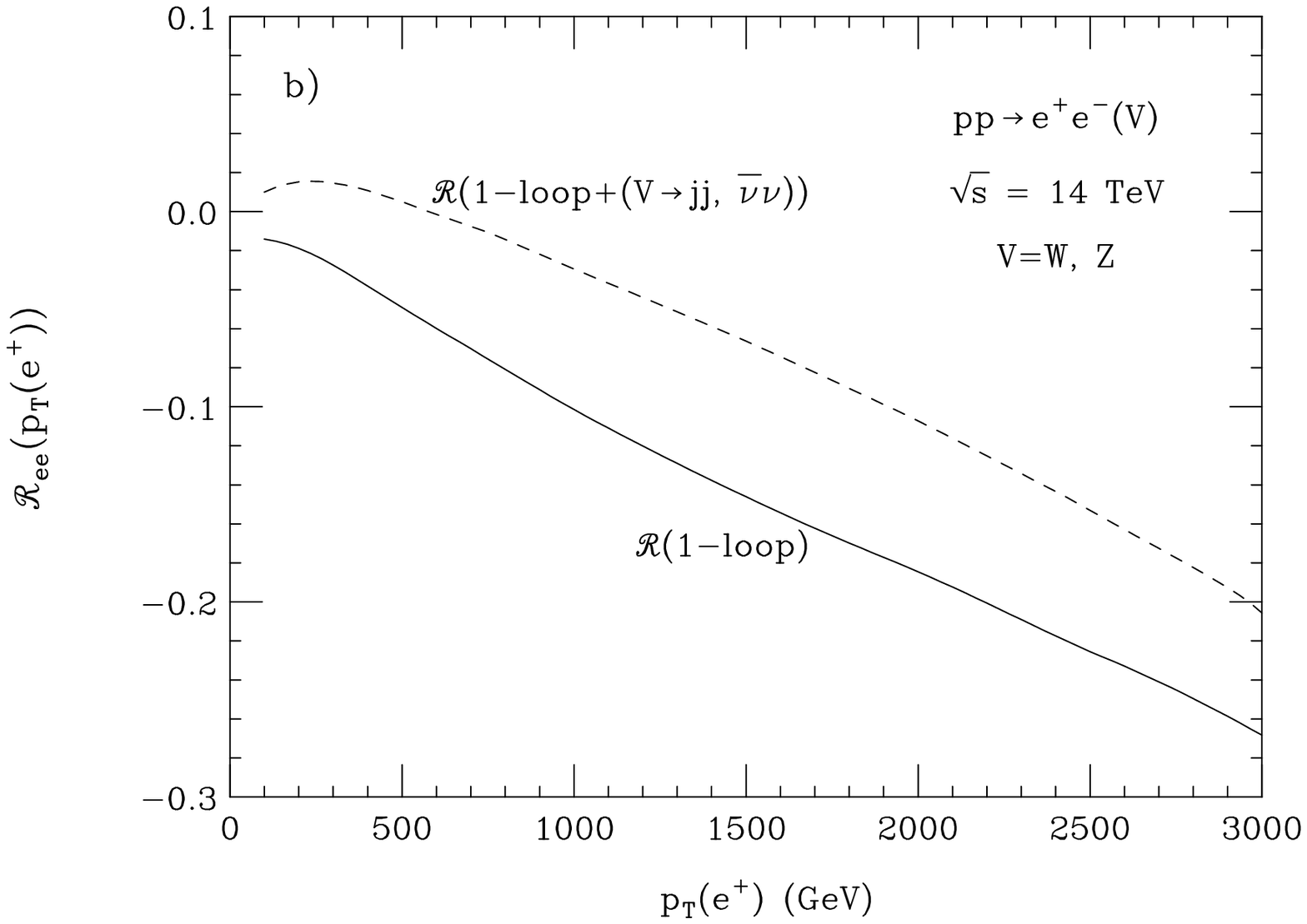}
\vspace*{2mm}
\caption[]{\label{fig:fig7} 
The relative correction with respect to the LO $e^+e^-$ cross section at the
LHC as a 
function a) of the $e^+e^-$ invariant mass and b) the positron $p_T$. 
The solid curve shows the result if only the ${\cal O}(\alpha)$
corrections of Ref.~\protect{\cite{Baur:2001ze}} are taken into
account. The 
dashed curve shows ${\cal R}_{ee}$ if ${\cal O}(\alpha^3)$
$e^+e^- V$ production with $V\to jj$ and $Z\to\bar\nu\nu$ is included as
well. The definition 
of ${\cal R}_{ee}$ and the cuts imposed are described in the text.}
\vspace{-7mm}
\end{center}
\end{figure}

Results for the neutral channel are shown in Fig.~\ref{fig:fig6}
for the Tevatron, and in Fig.~\ref{fig:fig7} for the LHC. 
To calculate the relative correction to the LO $e^+e^-$ cross section
which originate from the ${\cal O}(\alpha)$ weak one-loop and photonic
corrections, I have used the results of Ref.~\cite{Baur:2001ze}. The
one-loop weak corrections in neutral Drell-Yan production are seen to
have a smaller effect on the differential cross section than in the
charged channel. As in $e\nu$ production, weak boson emission effects in
the neutral channel are quite small at the Tevatron. They increase
${\cal R}_{ee}$ by less than~0.01 over 
most of the invariant mass and $p_T$ ranges considered. At the LHC, $W$
and $Z$ radiation increase ${\cal R}_{ee}$ by up to 0.06. For example,
at $m(e^+e^-)=2.0$~TeV, the relative correction to the differential
cross section without (with) weak boson emission is ${\cal
R}_{ee}(m(e^+e^-))= -0.108$ (${\cal R}_{ee}(m(e^+e^-))= -0.073$). For
comparison, the statistical uncertainty of the Drell-Yan cross section at
$m(e^+e^-)=2.0$~TeV is about 18\% for 100~fb$^{-1}$. $W$
and $Z$ radiation thus moderately reduce the size of the ${\cal
O}(\alpha)$ electroweak
radiative corrections to the neutral Drell-Yan cross section at the
LHC in the experimentally accessible invariant mass range.

The experimental and theoretical systematic uncertainties in charged and
neutral Drell-Yan production 
at the Tevatron are of ${\cal O}(10\%)$~\cite{wprime,zprime,grav}. A
similar result is expected at the
LHC~\cite{cmstdr,cmsdimuon}. Therefore, with the 
possible exception of neutral Drell-Yan production at the Tevatron,
electroweak radiative corrections and weak boson emission will have a
non-negligible effect.

\section{Di-boson Production}
\label{sec:sec4}

Di-boson production, $p\,p\hskip-7pt\hbox{$^{^{(\!-\!)}}$} \to
W^\pm\gamma,\, Z\gamma,\, W^\pm Z,\, ZZ,\, W^+W^-$, offers an
opportunity to probe the gauge boson
self-couplings~\cite{Ellison:1998uy}. At the LHC, 
$W^+W^-$ and $ZZ$ production are also of interest as background
processes to Higgs boson production~\cite{Buscher:2005re}. In order to
precisely measure the 
gauge boson self-couplings, accurate theoretical predictions are
needed. The NLO QCD corrections to di-boson production have been
calculated several years
ago~\cite{Frixione:1992pj,Baur:1993ir,Ohnemus:1994qp,Baur:1997kz,Baur:1994aj,nlodibos}.
More recently, the combined one-loop weak and photonic corrections to
these processes have been
computed~\cite{Accomando:2001fn,wgam,Accomando:2004de}. Contributions from
weak boson emission are not included in these calculations. Furthermore,
numerical results are presented only for the LHC. In the following, I
therefore consider di-boson production only at the LHC. 

The experimental systematic and the PDF uncertainties at the LHC for all
di-boson production processes are in the $5-15\%$ range. 
The uncertainty from higher order QCD corrections for the individual
processes is discussed in more detail below.

\subsection{$W\gamma$ and $Z\gamma$ production}

$W\gamma$ events are usually selected by requiring the $W$ boson to
decay leptonically.  For hadronic $W$ decays, QCD $\gamma jj$ production
constitutes a very large background. To identify $W\gamma$ events,
experiments therefore search for events with one isolated high
$p_T$ charged lepton, large missing transverse momentum, and an isolated
hard photon. To be specific, I impose the following cuts in the
calculation of the $W\gamma$ cross section at the LHC: 
\begin{eqnarray}
\label{eq:wgcut1}
p_T(\ell) >  25~{\rm GeV,} & \qquad & |\eta(\ell)| < 2.5, \\
\label{eq:wgcut2}
p_T(\gamma) >  50~{\rm GeV,} & \qquad & |\eta(\gamma)|  < 2.5,\\
\label{eq:wgcut3}
p\llap/_T  >  25~{\rm GeV,} & \qquad {\rm and} \qquad & \Delta
R(\ell\gamma)  > 0.4. 
\end{eqnarray}
Due to the relatively large photon $p_T$ cut, radiative $W$ decay,
$pp \to W\to\ell\nu\gamma$ is strongly
suppressed and henceforth will be ignored. To compute weak boson
emission effects in $W\gamma$ production, the cross sections for $pp\to
W\gamma V$ ($V=W^\pm,\,Z$) have to be calculated. 

\begin{figure}[t!] 
\begin{center}
\includegraphics[width=15.3cm]{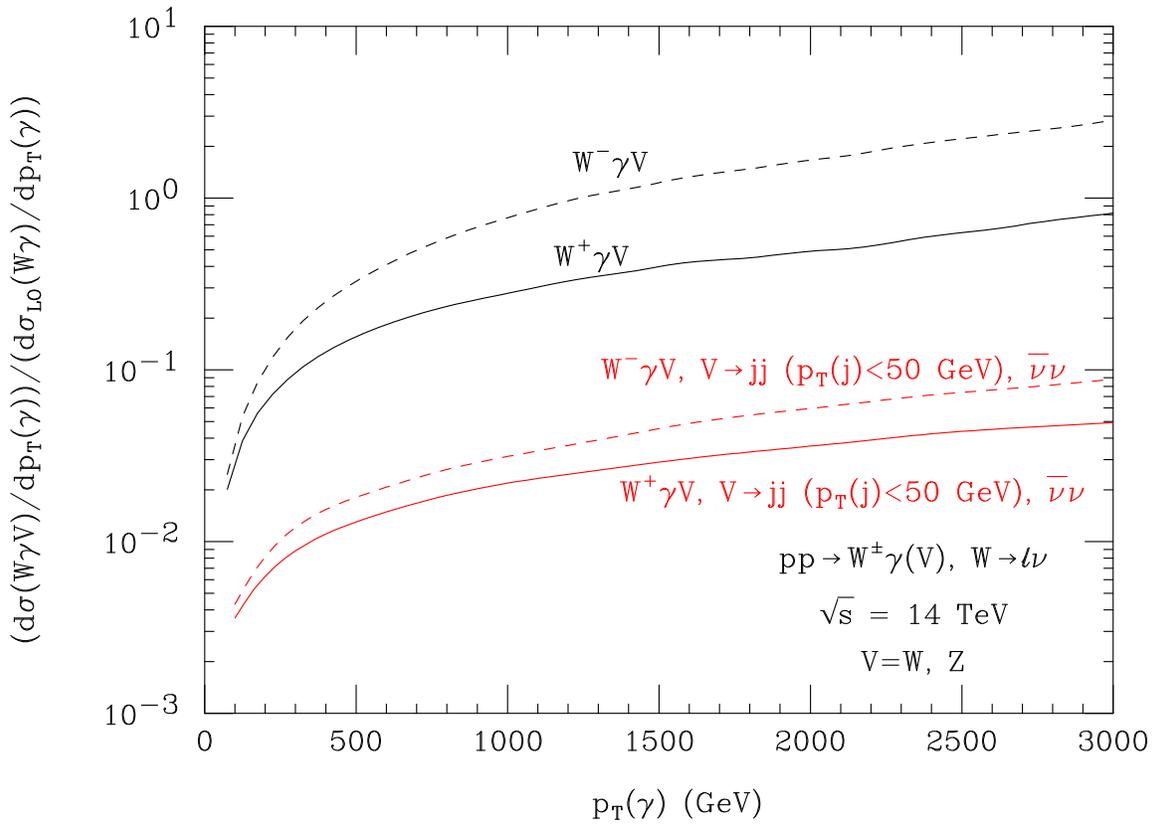} 
\vspace*{2mm}
\caption[]{\label{fig:fig8} 
Ratio of the $W\gamma V$ ($V=W^\pm,\,Z$) and the LO $W\gamma$ cross 
section as a function of the photon transverse
momentum at the LHC. The $W$ boson is required to decay leptonically. 
Results are shown for the inclusive case (black
solid and dashed lines), and for the case where jets with
$p_T(j)>50$~GeV and $|\eta(j)|<2.5$, and events with more than one
charged lepton, are vetoed (red solid and dashed lines). The cuts imposed
are listed in Eqs.~(\protect{\ref{eq:wgcut1}})
--~(\protect{\ref{eq:wgcut3}}).}  
\vspace{-7mm}
\end{center}
\end{figure}

Before discussing which $V$ decays should be considered,
it is instructive to consider the ratio of the $W\gamma V$ and the LO
${\cal O}(\alpha^2)$ $W\gamma$ cross section in the inclusive $V\to$~all
case. This ratio is shown as a function of the photon 
transverse momentum in Fig.~\ref{fig:fig8} (black solid and dashed
lines). Since anomalous $WW\gamma$ couplings lead to large deviations at high
values of photon $p_T$, the transverse momentum distribution of the
photon is of particular interest in $W\gamma$ production. With
100~fb$^{-1}$, $W\gamma$ events with a photon $p_T$ up to about 1~TeV
will be produced.
For inclusive $V$ decays, the $W\gamma V$ to $W\gamma$ cross section
ratio grows very quickly and, for large values of $p_T(\gamma)$, exceeds
the LO $W\gamma$ cross section. The effect is particularly pronounced in
the $W^-\gamma$ case. Naively one would expect that the $W\gamma V$
cross section is suppressed by ${\cal O}(\alpha)$ with respect to the LO
cross section, {\it ie.} the cross section ratio should be of ${\cal O}(0.1)$
or less. However, the LO $W\gamma$ cross section itself is suppressed by the
so-called ``radiation zero''~\cite{RAZ} which causes the $W\gamma V$ to
$W\gamma$ cross section ratio to be much larger than expected. For large
photon transverse momenta, the rate for $W^-\gamma$ production is
about a factor of four smaller than that for $W^+\gamma$
production. Since the $W\gamma V$ cross section is dominated by the
$W^+W^-\gamma$ channel, the $W\gamma V$ to $W\gamma$ cross section ratio is
larger in the $W^-\gamma$ channel.

The contributions of the weak boson emission processes
$pp \to W\gamma V$ to the ${\cal O}(\alpha^3)$ $W\gamma$ 
cross section have to be compared with
those of the combined ${\cal O}(\alpha)$ one-loop weak and photonic radiative
corrections~\cite{wgam}. The results are shown in
Table~\ref{tab:tab2}. 
\begin{table}[t!]
\caption{Relative size of the $W\gamma V$ contributions, $\delta(W\gamma
V)$, and the combined ${\cal
O}(\alpha)$ one-loop weak and photonic corrections, $\delta$(1-loop), 
to $W\gamma$ production at the LHC as a function of the
photon transverse momentum. Cross sections are normalized to the LO
$W\gamma$ cross section. The results for the ${\cal 
O}(\alpha)$ one-loop weak and photonic radiative corrections are taken from
Ref.~\protect\cite{wgam}. Results are shown for inclusive
$V$ decays ($\delta_{incl}$), and for the case where jets with
$p_T(j)>50$~GeV and $|\eta(j)|<2.5$, and events with more than one
charged lepton, are vetoed ($\delta_{veto}$). }
\label{tab:tab2}
\vskip 5.mm
\begin{tabular}{c|c|c|c|c|c}
$p_T(\gamma)$ & $\delta$(1-loop)~\cite{wgam} & $\delta_{incl}(W^+\gamma V)$ & 
$\delta_{incl}(W^-\gamma V)$  & $\delta_{veto}(W^+\gamma V)$  &
$\delta_{veto}(W^-\gamma V)$ \\
\tableline
275 GeV & -8.0\% & 9.4\% & 16.3\% & 0.6\% & 0.8\% \\
525 GeV & -17.0\%  & 17.0\%  & 33.8\%  & 1.5\%  & 2.1\%  \\
775 GeV & -23.4\%  & 22.8\%  & 56.5\%  & 1.9\%  & 2.7\% 
\end{tabular}
\end{table}
Note that Ref.~\cite{wgam} uses slightly different cuts and parameters
than I do. However, these effects should approximately cancel in
the cross section ratio. 
The relative sizes of the one-loop weak and photonic radiative
corrections for $W^+\gamma$ and $W^-\gamma$ production are approximately
equal. In $W^+\gamma$ production, $\delta$(1-loop) and
$\delta_{incl}(W^+\gamma V)$ approximately 
cancel for the range of photon transverse momenta listed here. On the
other hand, for $pp\to W^-\gamma$, a significant positive contribution 
remains when summing $\delta$(1-loop) and $\delta_{incl}(W^-\gamma V)$. 

As discussed above, experiments require one charged lepton in the selection of
$W\gamma$ events. Leptonic $V$ decays in $W\gamma V$ production
therefore have to be excluded, except for $Z\to\bar\nu\nu$. This will
reduce  $\delta_{incl}$ by about $20-30\%$.

Due to the suppression of the LO $W\gamma$ cross section and the
logarithmic growth of the $W\gamma j$ cross section with $p_T(\gamma)$
(see Eq.~(\ref{eq:nlo})), the NLO QCD corrections for $W\gamma$
production at the LHC are very large at high photon transverse
momenta~\cite{Baur:1993ir}. At large $p_T(\gamma)$, the NLO QCD $W\gamma$
cross section is dominated by the $qg\to W\gamma q'$ contribution, {\it
ie.} most $W\gamma$ events contain a hard jet. Since the NLO QCD corrections
significantly reduce the sensitivity to anomalous $WW\gamma$ couplings,
it is advantageous to impose a jet veto. This strongly reduces the
$W\gamma V$ cross section. We illustrate the impact of a jet veto in
Fig.~\ref{fig:fig8} and Table~\ref{tab:tab2} for the case where
jets with $p_T(j)>50$~GeV and $|\eta(j)|<2.5$ are vetoed, and only one
charged lepton in events is allowed. The jet veto
suppresses the $W\gamma V$ cross section by a factor $20-40$. About
one-half of the remaining $W\gamma V$ rate originates from $WZ\gamma$
production with $Z\to\bar\nu\nu$. Table~\ref{tab:tab2} shows that, if a
jet veto is imposed, the
contribution from weak boson emission processes to the ${\cal
O}(\alpha^3)$ cross section is much smaller than
that from one-loop weak and photonic radiative corrections. 

The results presented here demonstrate that contributions to the
$W\gamma$ cross section from weak boson emission may be as important as
those from the ${\cal O}(\alpha)$ virtual weak corrections. However,
they also show that the 
size of these contributions depends very strongly on the event selection
criteria. 

Since the NLO QCD $W\gamma$ cross section at high $p_T(\gamma)$ is
dominated by the tree level process $qg\to W\gamma q'$, it still 
depends considerably on the factorization and renormalization scales (see
eg. Ref.~\cite{Baur:1994aj}). The uncertainty 
from higher order QCD corrections in this region is roughly of the 
size of the contribution of the weak boson emission processes, {\it ie.}
in the $10-50\%$ range. Imposing a jet veto greatly reduces the scale
uncertainty. In this case, the full ${\cal O}(\alpha)$
electroweak corrections which include virtual corrections and weak boson
emission effects, will be 
significantly larger than the QCD scale uncertainty, and at least as
large as the combined PDF and experimental systematic uncertainties. 

$Z\gamma$, $Z\to\ell^+\ell^-$ events are selected by requiring two
isolated charged leptons with 
opposite electric charge, and a hard, isolated photon. In addition,
events should not have any significant amount of missing transverse
momentum. In the following calculation, I impose the lepton and photon
cuts of Eqs.~(\ref{eq:wgcut1}) --~(\ref{eq:wgcut3}), except the
$p\llap/_T$ cut. Instead, the missing transverse momentum in $Z\gamma$
events has to satisfy Eq.~(\ref{eq:veto}). The photon $p_T$ cut strongly
suppresses contributions from radiative $Z$ decays which I shall ignore
in the following.

\begin{figure}[t!] 
\begin{center}
\includegraphics[width=15.3cm]{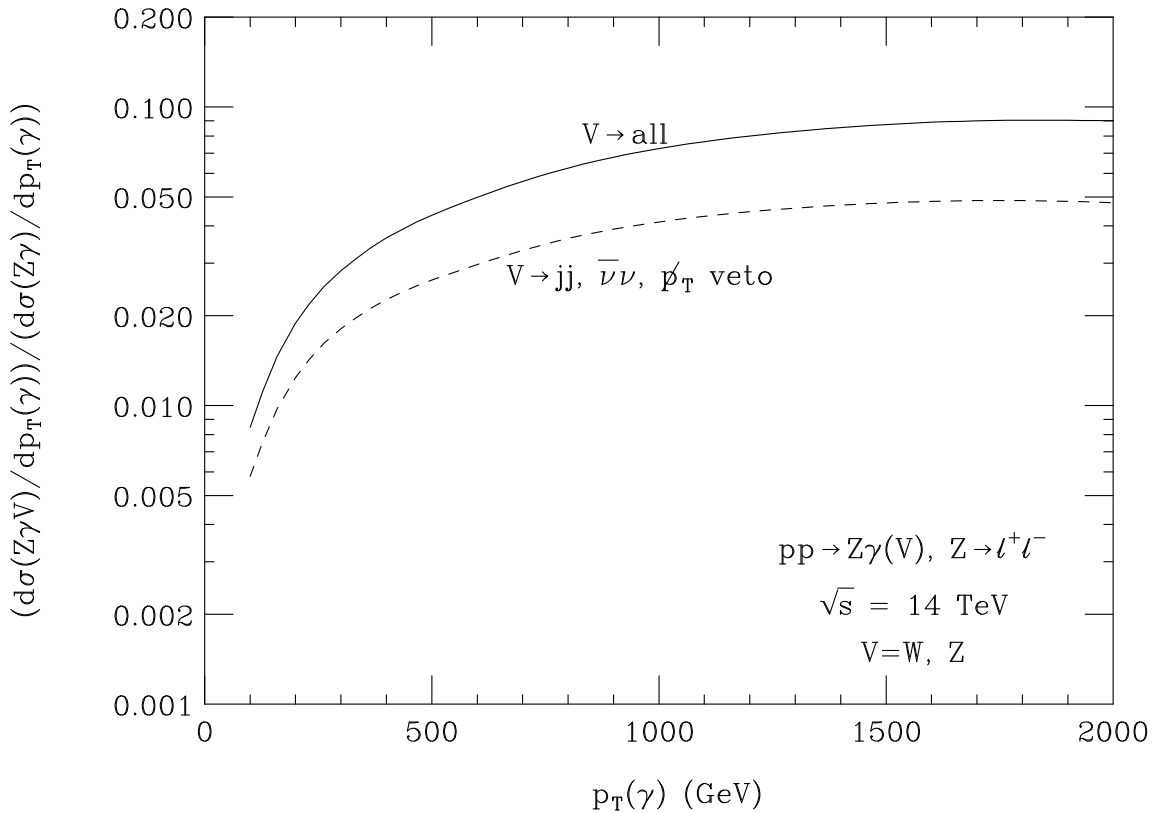} 
\vspace*{2mm}
\caption[]{\label{fig:fig9} 
Ratio of the $Z\gamma V$ ($V=W^\pm,\,Z$) and the LO $Z\gamma$ cross 
section as a function of the photon transverse
momentum at the LHC. The $Z$ boson is required to decay leptonically. 
Results are shown for the inclusive case, $V\to$~all
(solid line), and for the case where events with more than two
charged leptons and missing transverse momentum which does not satisfy
Eq.~(\ref{eq:veto}) are vetoed (dashed line). The cuts imposed are listed in
Eqs.~(\protect{\ref{eq:wgcut1}}) --~(\protect{\ref{eq:wgcut3}}).} 
\vspace{-7mm}
\end{center}
\end{figure}

In contrast to $W\gamma$ production there is no radiation zero in $pp\to
Z\gamma$. The cross section for $Z\gamma V$, $V\to$all, production
therefore is 10\% 
or less of the LO $Z\gamma$ rate over the entire photon transverse
momentum range. This is shown by the solid line in Fig.~\ref{fig:fig9}.
Since the LO $Z\gamma$ cross section is not suppressed, the NLO QCD
corrections, especially at high $p_T(\gamma)$, are much smaller than for
$W\gamma$ production, and there is no need to impose a jet veto when
analyzing anomalous couplings~\cite{Baur:1997kz}. Events with more than
two charged leptons, however, are not included in a $Z\gamma$
sample. Vetoing events with more than two charged leptons, imposing the
$p\llap/_T$ veto of Eq.~(\ref{eq:veto}), and requiring that the charged
leptons from $Z\to\ell^+\ell^-$ are isolated by
\begin{equation}
\label{eq:rlj}
\Delta R(\ell,j)>0.4
\end{equation}
from the jets which originate from $V$ decays in $Z\gamma V$ production,
one obtains the dashed line in Fig.~\ref{fig:fig9}. For
$p_T(\gamma)=300$~GeV (500~GeV), the $Z\gamma V$ cross section is
approximately 1.9\% (2.7\%) of the LO $Z\gamma$ cross section. For
comparison, the combined one-loop weak and photonic corrections,
normalized to the LO $Z\gamma$ rate, are $-15\pm 1\%$ ($-24\pm 2\%$) at
$p_T(\gamma)=300$~GeV (500~GeV)~\cite{wgam}. Weak boson emission effects
therefore only mildly affect the $Z\gamma$ production cross section. 

The uncertainties from higher order QCD corrections in $Z\gamma$
production are similar to those for $W\gamma$ production when a jet veto
is imposed. The full ${\cal O}(\alpha)$
electroweak corrections, including both virtual corrections and weak
boson emission effects, will be 
at least as large as the combined theoretical and experimental
systematic uncertainties. They cannot be neglected in a $Z\gamma$
analysis at the LHC.

\subsection{$WZ$ and $ZZ$ production}

Weak boson emission effects in $WZ$ and $ZZ$ production are very similar
to those in $pp\to W\gamma$ and $pp\to Z\gamma$, respectively. If one or
both of the weak bosons decay hadronically, the signal process is
swamped by QCD background. For $Z\to\bar\nu\nu$ and $W\to\ell\nu$, $WZ$
production cannot be discriminated from single $W$ production. To select
$WZ$ events one therefore requires three charged leptons and missing
transverse momentum. In $ZZ$ production, in order to
reduce the background sufficiently, either both $Z$
bosons have to decay into charged leptons, or one of them decays into
neutrinos and the other into charged leptons. In the following, I
concentrate on the 4~lepton final state in $ZZ$ production. To identify 
$WZ$ and $ZZ$ events, I impose the cuts listed in Eq.~(\ref{eq:wgcut1}). In
addition, in $WZ$ production, I require that the $p\llap/_T$ cut of
Eq.~(\ref{eq:wgcut3}) is satisfied. In $pp\to ZZ$, events which do not
satisfy Eq.~(\ref{eq:veto}) are rejected. 

\begin{figure}[t!] 
\begin{center}
\includegraphics[width=15.3cm]{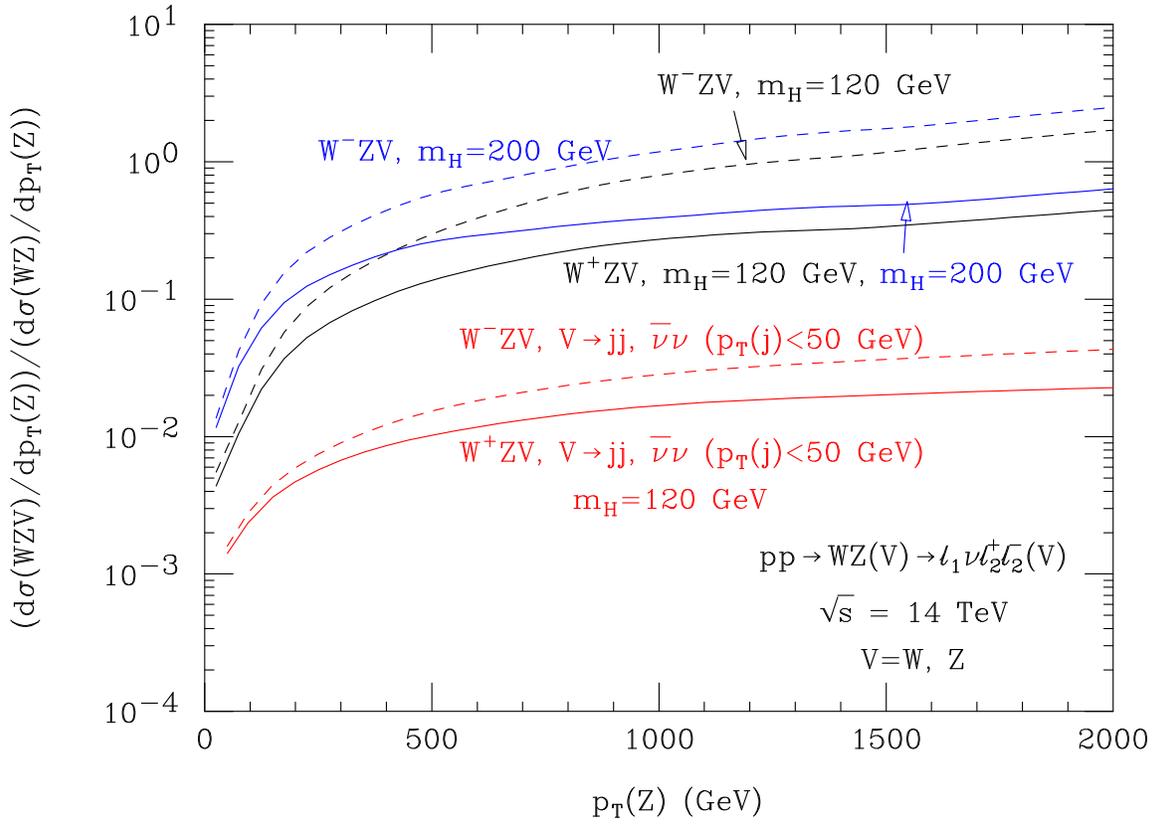}
\vspace*{2mm}
\caption[]{\label{fig:fig10} 
Ratio of the $WZV$ ($V=W^\pm,\,Z$) and the LO $WZ$ cross 
section as a function of the $Z$ transverse momentum at the LHC. Results
are shown for the inclusive case, $V\to$~all (black and blue solid and
dashed lines), and for the case where jets with
$p_T(j)>50$~GeV and $|\eta(j)|<2.5$, and events with more than three
charged leptons are vetoed (red solid and dashed lines). 
The black (blue) lines correspond to $m_H=120$~GeV ($m_H=200$~GeV). In the red
curves, the Higgs boson mass is fixed to $m_H=120$~GeV. 
The $W$ and $Z$ bosons are required to decay leptonically,
$WZ\to\ell_1\nu\ell_2^+\ell_2^-$ ($\ell_{1,2}=e,\,\mu$). 
The cuts imposed are discussed in the text.}
\vspace{-7mm}
\end{center}
\end{figure}

The LO $WZ$ cross section is suppressed by an approximate radiation
zero~\cite{Baur:1994ia}. There is no suppression mechanism in $ZZ$
production. It is therefore not surprising that the ratio of the $WZV$
and $WZ$ cross section rises quickly with $p_T(Z)$, and, for inclusive
$V$ decays, becomes ${\cal O}(1)$
in the TeV region. This is shown by the black and blue lines 
in Fig.~\ref{fig:fig10}. The transverse momentum distribution of the $Z$
boson is of particular interest because of its sensitivity to anomalous
$WWZ$ couplings. For 100~fb$^{-1}$, $WZ$ events with a $Z$-boson $p_T$
up to about 500~GeV will be produced.
The weak boson emission processes contributing to $WZ$ and $ZZ$
production, $pp\to WZV$ and $pp\to ZZV$ involve Higgs exchange
diagrams. The relative cross section thus depends on the Higgs boson
mass, $m_H$. The black lines in Fig.~\ref{fig:fig10} correspond to
$m_H=120$~GeV which is close to the lower limit established by
LEP2~\cite{Barate:2003sz}. The blue curves show the results for
$m_H=200$~GeV, the current upper 95\%~CL limit from a fit to all electroweak
data~\cite{moscow}. The cross sections for $WZV$ and $ZZV$
production vary significantly with $m_H$ only for small values of
$p_T(Z)$. At large transverse momenta, $V$ bremsstrahlung diagrams
dominate, and the cross section depends only slightly on the Higgs boson
mass. Since the $W^-Z$ cross section is significantly smaller than that
for $W^+Z$ production, and the $WZV$ cross section is dominated by
$W^+W^-Z$ production, the cross section ratio is larger in the $W^-Z$
case. 

As in $W\gamma$ production, the NLO QCD corrections to $pp\to WZ$ become
very large in the high $p_T(Z)$ region, and it is
advantageous to impose a jet veto~\cite{Baur:1994aj}. Requiring that
there are no jets with $p_T(j)>50$~GeV and $|\eta(j)<2.5$ in the event
reduces the $WZV$ cross section to a few percent or less of the LO $WZ$
rate. This is shown for $m_H=120$~GeV by the red solid dashed lines in
Fig.~\ref{fig:fig10}. In order not to overburden the figure, only results
for $m_H=120$~GeV are shown when a jet veto is imposed. The Higgs
mass dependence with a jet veto imposed is similar to that encountered
in the inclusive case.

The relative rate for weak boson emission in $WZ$ production should be
compared with that of the ${\cal O}(\alpha)$ virtual weak
corrections. The combined 
one-loop weak and photonic corrections to $WZ$ production were
calculated in the high energy limit in Ref.~\cite{Accomando:2004de} and
listed as a function of the minimum transverse momentum of
the $Z$ boson, $p_T^{min}(Z)$. These results are compared with the
relative rate for 
weak boson emission in Table~\ref{tab:tab3}. The relative rate $\Delta$
shown in the Table is defined by
\begin{equation}
\Delta(X)=
{\sigma^X(p_T(Z)>p_T^{min}(Z))\over\sigma_{LO}^{WZ}(p_T(Z)>p_T^{min}(Z))
}.
\end{equation}
\begin{table}[t!]
\caption[]{Relative size of the $WZV$ contributions, $\Delta(WZV)$, and
the combined ${\cal 
O}(\alpha^3)$ one-loop weak and photonic corrections, $\Delta$($WZ$, 1-loop), 
to $WZ$ production at the LHC as a function of the minimum $Z$ boson transverse
momentum, $p_T^{min}(Z)$. Cross sections are normalized to the LO $WZ$
cross section. The results for the combined ${\cal 
O}(\alpha^3)$ one-loop weak and photonic radiative corrections are taken from
Ref.~\protect\cite{Accomando:2004de}. Results are shown for inclusive
$V$ decays ($\Delta_{incl}$), and for the case where jets with
$p_T(j)>50$~GeV and $|\eta(j)|<2.5$, and events with more than three
charged lepton are vetoed ($\Delta_{veto}$). }
\label{tab:tab3}
\vskip 5.mm
\begin{tabular}{cdddd}
$p_T^{min}(Z)$ & 250~GeV & 300~GeV & 400~GeV & 500~GeV\\
\tableline
$\Delta$($WZ$, 1-loop)~\cite{Accomando:2004de} & $-$10.9\% & $-$13.1\% &
$-$17.8\% & $-$21.2\% \\
\tableline
$\Delta_{incl}(W^+ZV)$, $m_H=120$~GeV & 9.7\% & 11.1\% & 15.0\% & 17.7\% \\
$\Delta_{incl}(W^-ZV)$, $m_H=120$~GeV & 18.1\% & 21.7\% & 31.9\% & 41.7\% \\
$\Delta_{veto}(W^+ZV)$, $m_H=120$~GeV  & 0.8\% & 0.9\% & 1.1\% & 1.3\% \\
$\Delta_{veto}(W^-ZV)$, $m_H=120$~GeV  & 1.1\% & 1.2\% & 1.5\% & 1.8\%\\
\tableline
$\Delta_{incl}(W^+ZV)$, $m_H=200$~GeV & 18.7\% & 24.4\% & 29.4\% &
31.3\% \\
$\Delta_{incl}(W^-ZV)$, $m_H=200$~GeV & 35.8\% & 45.5\% & 65.1\% &
76.8\% \\
$\Delta_{veto}(W^+ZV)$, $m_H=200$~GeV  & 1.5\% & 2.0\% & 2.1\% & 2.2\%
\\
$\Delta_{veto}(W^-ZV)$, $m_H=200$~GeV  & 2.1\% & 2.6\% & 3.0\% & 3.4\%
\end{tabular}
\end{table}
Without a jet veto, weak boson emission effects are as large as or
larger than the combined virtual weak and photonic corrections to $WZ$
production. If a jet veto is imposed, they become small. The calculation of
$\Delta$($WZ$, 1-loop)~\cite{Accomando:2004de} uses a 
slightly different rapidity cut on the leptons. Furthermore, in order to
ensure the validity of the high energy approximation, a cut on the
rapidity difference between the $Z$ boson and the lepton from
$W\to\ell\nu$ of $\Delta y(Zl)<3$ was imposed. $\Delta_{incl}$ and
$\Delta_{veto}$ were calculated with the lepton rapidity cut of
Eq.~(\ref{eq:wgcut1}) and without a $\Delta y(Zl)$ cut. The dependence
of the cross section ratios $\Delta_{incl}$ and $\Delta_{veto}$ on these
cuts should, however, be mild. Nevertheless, this should be
kept in mind when comparing the numbers for $\Delta$($WZ$, 1-loop),
$\Delta_{incl}$ and $\Delta_{veto}$ in Table~\ref{tab:tab3}.

\begin{figure}[t!] 
\begin{center}
\includegraphics[width=15.3cm]{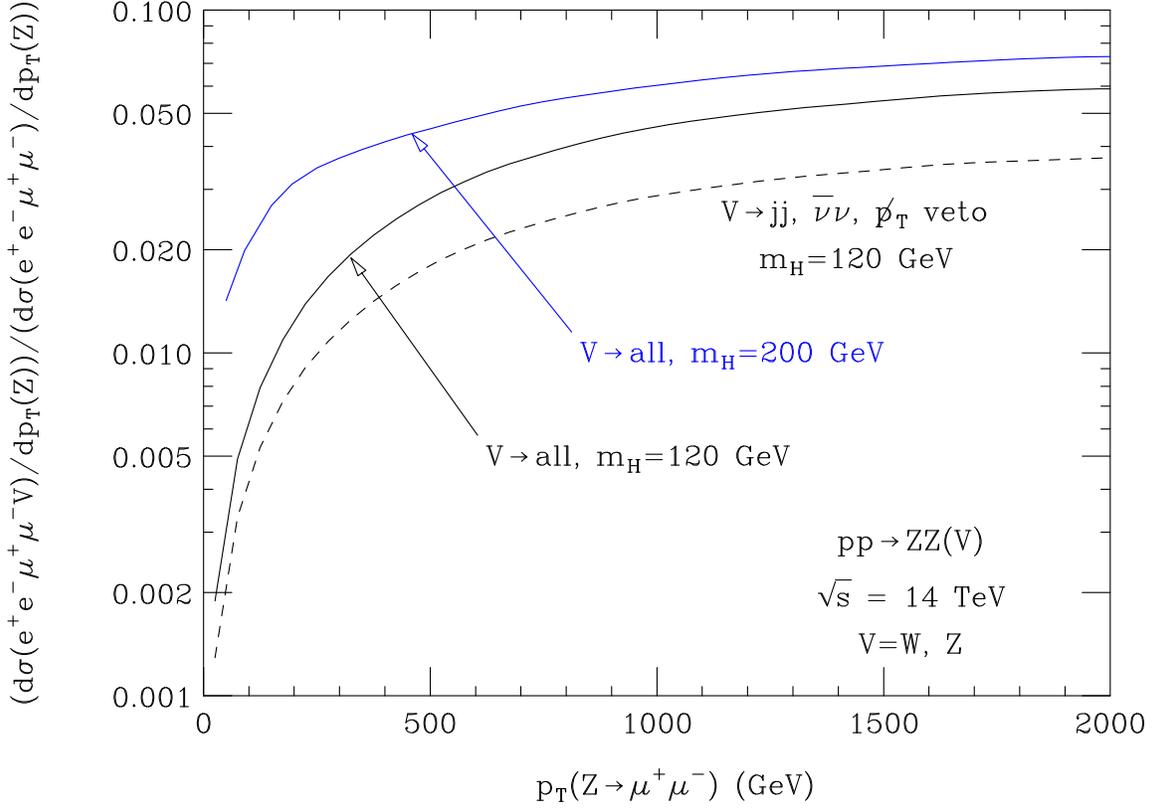}
\vspace*{2mm}
\caption[]{\label{fig:fig11} 
Ratio of the $ZZV$ and the LO $ZZ$ cross section
as a function of $p_T(Z\to\mu^+\mu^-)$ at the LHC. Only the $ZZ\to
e^+e^-\mu^+\mu^-$ final state is considered. 
Results are shown for the inclusive case, $V\to$~all
(black and blue solid lines), and for the case where events
with leptonic decays of the third weak boson, $V$, are not allowed and a
$p\llap/_T$ veto is imposed (dashed line). 
The black (blue) lines correspond to $m_H=120$~GeV ($m_H=200$~GeV). 
The cuts imposed are discussed in the text.}
\vspace{-7mm}
\end{center}
\end{figure}

In $ZZ$ production, the LO cross section is not suppressed, and weak
boson effects are of ${\cal O}(10\%)$ or less. This is shown in
Fig.~\ref{fig:fig11}, where the $ZZV$ cross section, normalized to the
LO $ZZ$ rate, is displayed as a function of $p_T(Z)$  for the $ZZ\to
e^+e^-\mu^+\mu^-$ channel. 
The black and blue solid lines show results for $m_H=120$~GeV and
$m_H=200$~GeV for the inclusive $V\to$~all case. The dashed line represents
the cross section ratio when events with more than four charged leptons
are rejected, charged leptons are required to be isolated, $\Delta
R(\ell,j)>0.4$, and the $p\llap/_T$ veto of Eq.~(\ref{eq:veto}) is 
imposed. The combined virtual weak and photonic corrections to $ZZ$
production in the high energy approximation increase from about $-20\%$
for $p_T(Z)=300$~GeV to $\approx -50\%$ 
at $p_T(Z)=900$~GeV~\cite{Accomando:2004de}. As in the $Z\gamma$ case,
weak boson emission effects in $ZZ$ production are substantially smaller
than the ${\cal O}(\alpha)$ virtual weak radiative corrections.

The uncertainties from higher order QCD corrections in $WZ$ ($ZZ$)
production are similar to those encountered in $W\gamma$ ($Z\gamma$)
production. The ${\cal O}(\alpha)$ electroweak corrections to $WZ$ and
$ZZ$ production, combining the virtual corrections in the high energy
approximation and 
weak boson emission effects, will thus be significantly larger than the
theoretical and experimental systematic uncertainties over most
of the $Z$ boson transverse momentum range. 

\subsection{$WW$ Production}

\begin{figure}[t!] 
\begin{center}
\includegraphics[width=15.3cm]{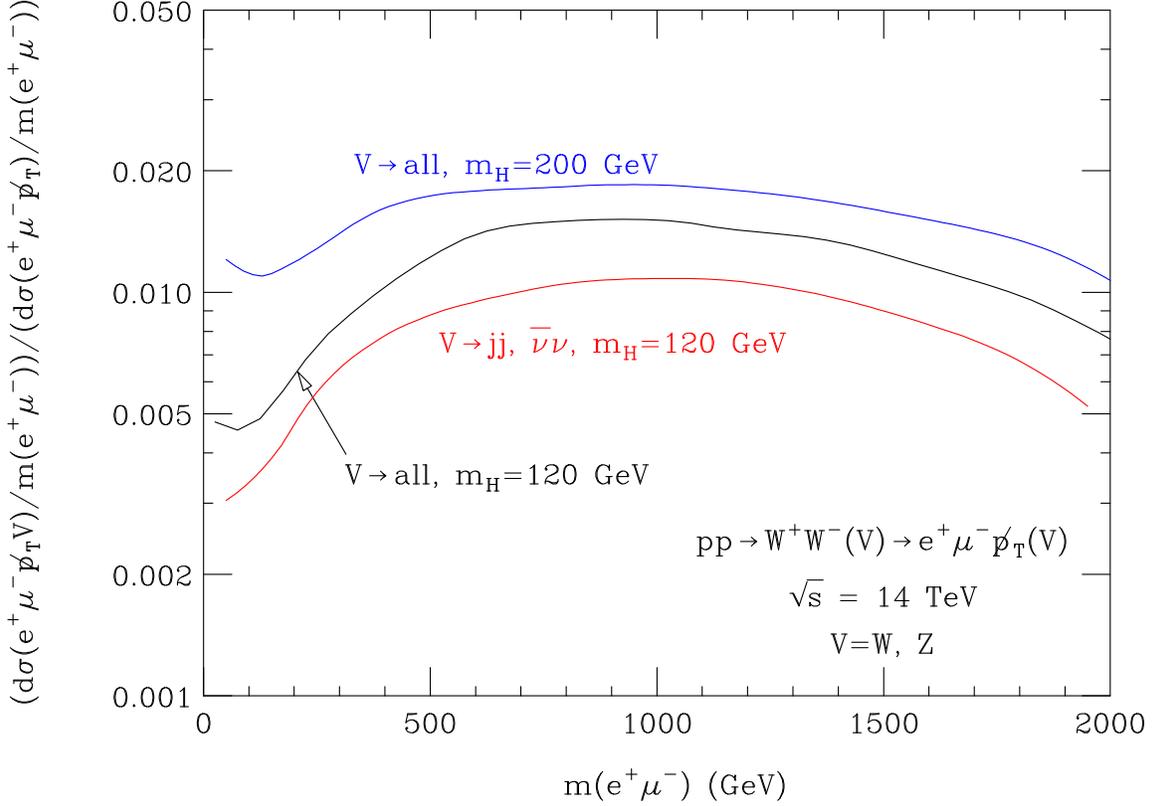}
\vspace*{2mm}
\caption[]{\label{fig:fig12} 
Ratio of the $WWV$ and the LO $WW$ cross section as a function of the
invariant mass of the two charged leptons, $m(e^+\mu^-)$, for the $W^+W^-\to 
e^+\mu^-p\llap/_T$ final state at the LHC.
Results are shown for the inclusive case, $V\to$~all
(black and blue lines), and for the case where events
with leptonic decays of the third weak boson, $V$, are not allowed
(red line). No restrictions on the jet activity in events are
imposed. 
The black (blue) lines correspond to $m_H=120$~GeV ($m_H=200$~GeV). 
The cuts imposed are discussed in the text.}
\vspace{-7mm}
\end{center}
\end{figure}

$W$ pair events are selected by requiring that both $W$ bosons decay
leptonically. In addition to two isolated charged leptons, a cut on the missing
transverse momentum is imposed. In the calculations presented in this
Section, I impose the cuts of Eqs.~(\ref{eq:wgcut1}) 
and~(\ref{eq:wgcut3}) and concentrate on the $W^+W^-\to
e^+\mu^-p\llap/_T$ final state. Since there is no radiation zero present
in the $q\bar q\to W^+W^-$ helicity amplitudes, one would naively expect
that the weak boson emission processes $pp\to WWV$ contributions are
relatively small, as in the $Z\gamma$ and $ZZ$ cases. However, this is
only true in some kinematic distributions, such as the invariant mass
distributions of the two leptons which is shown in Fig.~\ref{fig:fig12}.
The $WWV$ to LO $WW$ cross section ratio as a function of the invariant
mass of the two leptons is seen to be of ${\cal O}(1\%)$ or less once
events with three or more leptons have been eliminated (red line). No
restrictions on the jet activity, except for a $\Delta R(\ell,j)>0.4$
cut, are imposed in results shown in 
Fig.~\ref{fig:fig12}. A jet veto, which is advantageous in suppressing the
$t\bar t$ background~\cite{Acosta:2005mu}, would considerably reduce the cross
section ratio. The combined virtual weak and photonic ${\cal O}(\alpha)$
corrections to $W$-pair production in the high energy approximation
reduce the LO $WW$ cross section by 
$14\%$ for $m(e^+\mu^-)\geq 500$~GeV and $23\%$ for $m(e^+\mu^-)\geq
1$~TeV~\cite{Accomando:2004de}. Weak boson emission thus plays a minor
role for the di-lepton 
invariant mass distribution in $WW$ production.

The situation is completely different for the transverse momentum distribution
of the di-lepton system, which is particularly sensitive to anomalous
$WWV$ couplings~\cite{Baur:1995uv}. In the SM, the dominant $W^\pm$ helicity at
high energies in 
$\bar uu\to W^+W^-$ ($\bar dd\to W^+W^-$) is $\lambda_{W^\pm}=\mp 1$
($\lambda_{W^\pm}=\pm 1$)~\cite{LAGRANGIAN,BS,HEL} because of a
$t$-channel pole factor which peaks at small scattering angles with an
enhancement factor which is proportional to $\hat s$. Due to the $V-A$
nature of the $W\ell\nu$ coupling, the angular
distribution of the charged lepton in the rest frame of the parent $W$ is
proportional to $(1+Q_W\lambda_W\cos\theta)^2$, where $Q_W$ is the
$W$ charge and $\theta$ is the
angle with respect to the flight direction of the $W$ in the parton
center of mass frame. As a result, the charged leptons tend
to be emitted either both into ($\bar dd$ annihilation), or both
against the flight direction of their parent $W$ boson ($\bar uu$
annihilation), {\it i.e.}, they reflect the kinematic properties of the
$W$ bosons. At leading order, the $W^+$ and the $W^-$ in $W$ pair
production are back to back in the transverse plane, and the
transverse momenta  of the two leptons tend to cancel at high
energies. Above the $W$ threshold, the SM $p_T(e^+\mu^-)$ 
distribution thus drops very rapidly. 

\begin{figure}[t!] 
\begin{center}
\includegraphics[width=15.3cm]{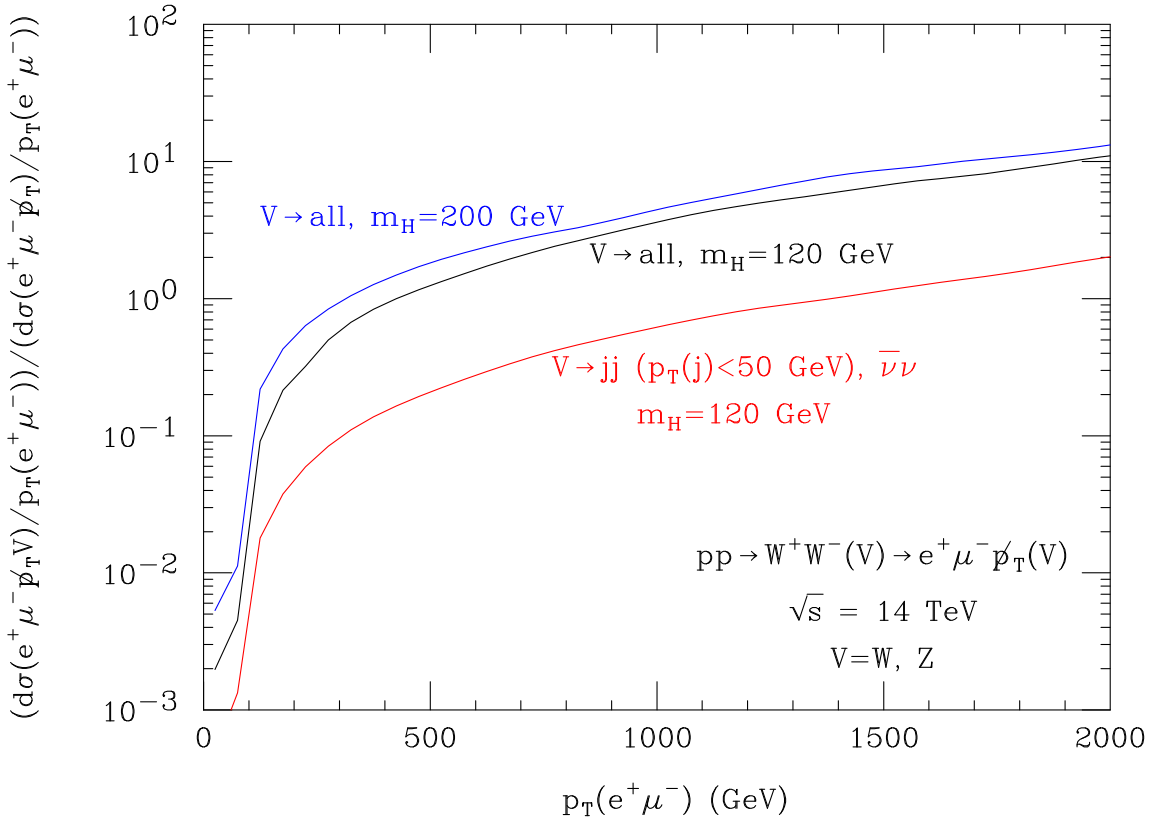}
\vspace*{2mm}
\caption[]{\label{fig:fig13} 
Ratio of the $WWV$ and the LO $WW$ cross section as a function of the
transverse momentum of the two charged leptons, $p_T(e^+\mu^-)$, for the
$W^+W^-\to e^+\mu^-p\llap/_T$ final state at the LHC.
Results are shown for the inclusive case, $V\to$~all
(black and blue lines), and for the case where events
with leptonic decays of the third weak boson, $V$, are not allowed
and jets with $p_T(j)>50$~GeV and $|\eta(j)|<2.5$ are vetoed (red line). 
The black (blue) line corresponds to $m_H=120$~GeV ($m_H=200$~GeV). The
Higgs boson mass is taken to be $m_H=120$~GeV in the red line.
The cuts imposed are discussed in the text.}
\vspace{-7mm}
\end{center}
\end{figure}

The delicate balance of the lepton transverse momenta, however, is
spoiled by real emission processes such as $pp\to WWV$. At
large transverse momenta, weak boson emission therefore affects the
$p_T(e^+\mu^-)$ differential cross section much more than other
distributions. This is evident in Fig.~\ref{fig:fig13} where I show the
$WWV$ to LO $WW$ cross section ratio as a function of the di-lepton
transverse momentum. 
In the inclusive case, $V\to$~all, the $WWV$ rate exceeds the LO $WW$ cross
section for $p_T(e^+\mu^-)>400$~GeV. The balance of the lepton
transverse momenta is also upset by gluon
radiation~\cite{Baur:1995uv}. A jet veto helps reducing the size of the
QCD corrections in the $p_T(e^+\mu^-)$ distribution. However, even when
a jet veto is imposed, the $WWV$ to LO $WW$ cross section ratio still
reaches about 15\% at $p_T(e^+\mu^-)=400$~GeV, the maximum di-lepton
transverse momentum which can be probed at the LHC with an integrated
luminosity of 300~fb$^{-1}$.

Ref.~\cite{Accomando:2004de} does not give results for how virtual
weak and photonic radiative corrections in the high energy approximation
affect the di-lepton transverse momentum distribution. Nevertheless, it
is possible to make a qualitative statement about their size. At high
energies, the virtual weak corrections are substantially larger than the
photonic corrections. For $2\to 2$ $W$ pair production, the $p\llap/_T$
and $p_T(e^+\mu^-)$ distributions are equal in absence of detector
effects. In the high energy regime, the virtual weak corrections thus
have a similar effect on the $p\llap/_T$ and the di-lepton $p_T$
distribution. For $p\llap/_T>200$~GeV, the combined virtual weak and
photonic ${\cal O}(\alpha)$ corrections reduce the $WW$ cross section by
about 20\%~\cite{Accomando:2004de}. It is therefore expected 
that the virtual weak corrections and the weak boson emission effects are
roughly of the same magnitude and partially cancel even when a jet veto is
imposed. Without a jet veto, $W$ and $Z$ radiation dominates over the
virtual weak corrections for $p_T(e^+\mu^-)$ larger than about 200~GeV.

Uncertainties from higher order QCD corrections in $WW$ production at
the LHC are similar in size to those encountered for the other di-boson
production processes. For the di-lepton invariant mass distribution, the
combined ${\cal O}(\alpha)$ virtual weak corrections and weak boson
emission effects thus will be as
large or larger than the combined theoretical and experimental
systematic uncertainties. Although weak boson emission significantly
reduces the size of the ${\cal O}(\alpha)$ electroweak corrections in
the di-lepton $p_T$ distribution, they are still non-negligible when
compared with the expected systematic uncertainties.

\section{Top Quark Production}
\label{sec:sec5}

In this Section, I investigate weak boson emission in top quark
production processes. I consider top pair production, $t$-channel single top
production, and $tW$ production. There are three different types of
single top quark production which can be distinguished by the virtuality
of the $W$ boson exchanged. In $s$-channel single top production, $q\bar
q'\to W^*\to t\bar b,\, \bar tb$, the squared four-momentum of the $W$
is positive, $Q_W^2>0$. In $t$-channel single top production, the $W$ is
exchanged in the $t$-channel and $Q_W^2<0$. Finally, in $tW$
production, the $W$ is on-shell, $Q_W^2=M_W^2$. Weak boson emission in
$s$-channel single top production has been studied in
Ref.~\cite{Ciafaloni:2006qu}. $t\bar bW$ production receives a large
contribution from ${\cal O}(\alpha_s^2)$ $\bar tt$ production and has
been found to be one of the dominant background sources
for $s$-channel single top production~\cite{Stelzer:1998ni}. $s$-channel
single top production therefore is not considered here. 

The top quark mass
used in all calculations here is $m_t=173$~GeV. This value agrees,
within errors, with the most recent world
average~\cite{mtop}. $b$-tagging efficiencies are not included in any 
numerical results presented in this Section. The ${\cal O}(\alpha)$
electroweak radiative corrections to 
$\bar tt$ production at the Tevatron have been found to be quite
small~\cite{Bernreuther:2006vg}. For $t$-channel single top and $tW$
production, they have only been calculated for the
LHC~\cite{Beccaria:2006ir,Beccaria:2006dt}. In the following, I
therefore concentrate on top quark production at the LHC.

\subsection{$\bar tt$ Production}

Top quark pair production at hadron colliders is important for several
reasons. The top quark mass is a fundamental parameter of the SM and
therefore should be measured as precisely as
possible~\cite{mtop}. Measuring the $\bar tt$ cross section provides a
test of the top quark production mechanism~\cite{doug}. At the LHC,
$pp\to\bar tt$ is also an important background to Higgs boson
production~\cite{Kinnunen:1999ak}. 

In the following, I only consider 
the $\ell\nu b\bar b+$jets final state; {\it ie.} one top
quark is required to decay semi-leptonically, and the other
hadronically. Both $b$-quarks in the final state are assumed to be
tagged. The following cuts are imposed on the
final state particles:
\begin{eqnarray}
p_T(\ell)>20~{\rm GeV}, & \qquad & |\eta(\ell)|<2.5, \\
p_T(j)>30~{\rm GeV}, & \qquad & |\eta(j)|<2.5, \\
p_T(b)>30~{\rm GeV}, & \qquad & |y(b)|<2.5, \\ \label{eq:dr}
p\llap/_T>40~{\rm GeV}, & \qquad & \Delta R(i,k)>0.4,
\end{eqnarray}
with $i,\,k=\ell,\,b,\,j$ and $i\neq k$. The number of non-tagged
isolated jets
in the event, $n(j)$, is required to be $n(j)\geq 2$; a jet veto, {\it
ie.} requiring 
$n(j)=2$, usually is not imposed~\cite{cmstdr,Beneke:2000hk}. One
combination of a tagged $b$-quark and two isolated jets has to be
consistent with originating from a hadronically decaying top quark. 
Only one charged lepton is 
allowed in the event. These requirements suppress $\bar ttV$ production
with $V=W\to\ell\nu$ and $V=Z\to\ell^+\ell^-$ and I will ignore these
channels here. However, $pp\to\bar ttV$ with $V\to jj$ and
$V=Z\to\bar\nu\nu$ have to be considered when calculating the
contribution of weak boson emission processes to top pair
production. Finally, in order to suppress $\bar ttW$ production with
$\bar tt\to\bar bb+4$~jets and $W\to\ell\nu$, I require  
\begin{equation}
\label{eq:cuts4}
\chi^2_{min}=\min_{b_1j_1j_2b_2j_3j_4~perm}
\left[\chi^2(b_1j_1j_2;b_2j_3j_4)\right] >4,
\end{equation}
where $\chi^2_{min}$ is the minimum of the $\chi^2(b_1j_1j_2;b_2j_3j_4)$
values of all possible combinations of jet pairs and $bjj$
combinations, and
\begin{eqnarray}
\chi^2(b_1j_1j_2;b_2j_3j_4)& = &{(m(j_1j_2)-M_W)^2\over\sigma_W^2}+
{(m(j_3j_4)-M_W)^2\over\sigma_W^2}+\\ \nonumber
& & {(m(b_1j_1j_2)-m_t)^2\over\sigma_t^2}
+{(m(b_2j_3j_4)-m_t)^2\over\sigma_t^2}.
\end{eqnarray}
For the $W\to jj$ and $t\to bjj$ mass resolutions I assume
$\sigma_W=7.8$~GeV and $\sigma_t=13.4$~GeV~\cite{Beneke:2000hk}. Only
$t$ and $\bar t$ resonant diagrams are included in the calculation. Diagrams
where the $W$ or $Z$ boson is emitted from one of the $t$ or $\bar t$ decay
products are not taken into account.

\begin{figure}[t!] 
\begin{center}
\includegraphics[width=15.3cm]{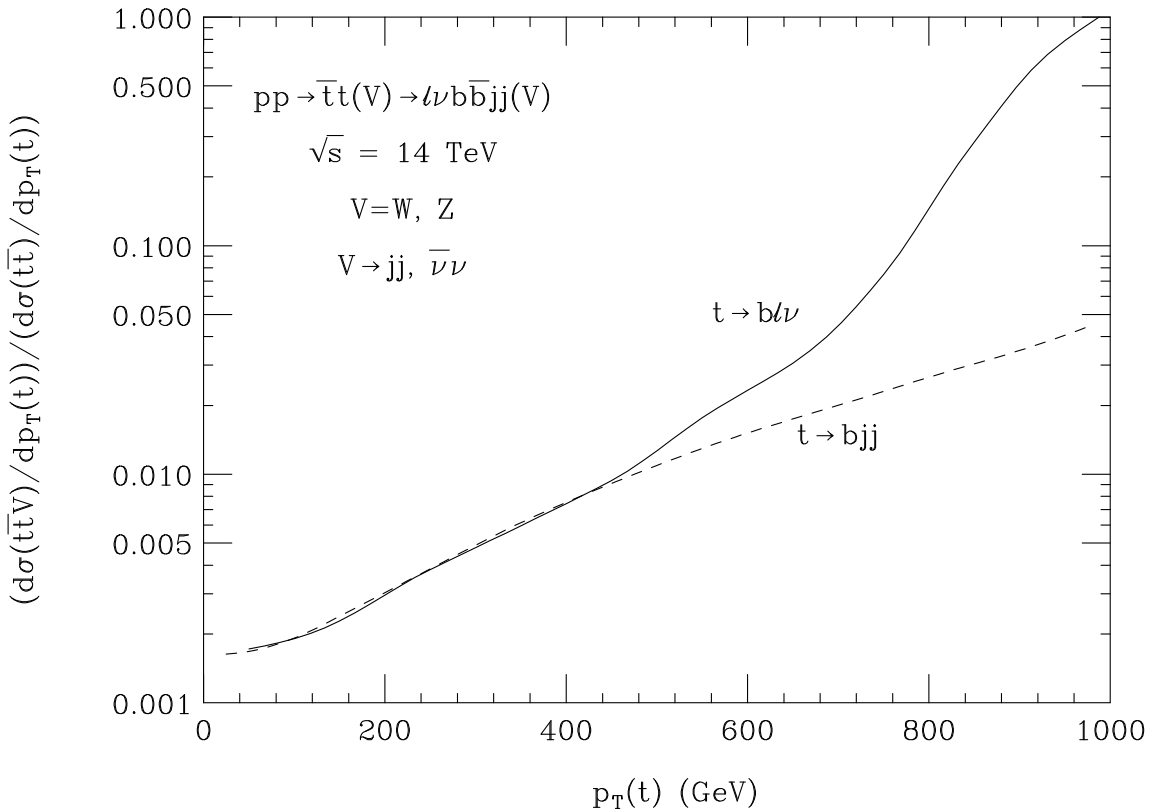}
\vspace*{2mm}
\caption[]{\label{fig:fig14} 
Ratio of the $\bar ttV$ and the LO $\bar tt$ cross section as a function
of the 
transverse momentum of the $t$ quark. The solid (dashed) curve
shows the result for $t\to b\ell\nu$ ($t\to bjj$).
The cuts imposed are discussed in the text.}
\vspace{-7mm}
\end{center}
\end{figure}

The $\bar ttV$ to LO $\bar tt$ cross section ratio as a function of the
transverse momentum of the $t$ quark is shown in
Fig.~\ref{fig:fig14}. The
transverse momentum distributions for $\bar t\to\bar b\ell\nu$ and $\bar
t\to\bar bjj$ are
equal to those for semileptonic and hadronic $t$ decays and
therefore are not shown. For the cuts imposed, and with an integrated
luminosity of 300~fb$^{-1}$, it should be possible to observe top quarks
from $\bar tt$ production  with a transverse momentum of up to 1~TeV at
the LHC. 
Since gluon fusion does not contribute to $\bar ttW$ production, the
inclusive $\bar ttV$ cross section is dominated by $pp\to\bar ttZ$.
Below a transverse momentum of about
400~GeV, the cross section ratios for semileptonic ($t\to b\ell\nu$) and
hadronic top decays ($t\to bjj$) are essentially identical. For larger
values of $p_T$, the cross section ratio for $t\to b\ell\nu$ grows
significantly faster.
The different behavior for semileptonic and hadronic top quark decays
for $p_T(t)>400$~GeV can be traced to the
separation cut imposed on the final state particles (see
Eq.~(\ref{eq:dr})). The separation cut is crucial for identifying $\bar
tt\to\ell\nu \bar bbjj$ events. 
For top quarks with very high transverse momenta,
the decay products are highly boosted and thus almost collinear. This
makes it increasingly difficult to satisfy Eq.~(\ref{eq:dr}). Since
there is no separation cut imposed on the neutrino in $t\to
b\ell\nu$, the $\Delta R$ cut is affecting the decay $t\to b
jj$ more significantly. 

Because the separation cut suppresses hadronic $t$ and $\bar t$ decays at high
transverse momentum, the top quark transverse momentum in $\bar ttV$
production is balanced by the $p_T$ of the vector 
boson $V$ for $t\to b\ell\nu$ and $p_T(t)>500$~GeV. The transverse
momentum of the hadronically decaying $\bar t$ is typically small. 
On the other hand, in LO $pp\to\bar tt$, $p_T(t)=p_T(\bar
t)$. This implies that the growth of the $\bar ttV$ to LO $\bar tt$ cross
section ratio at large $p_T(t)$ for semileptonic top decays is largely
due to kinematic effects, and not a result of soft and/or collinear weak boson
emission. The cross section ratio for $t\to b\ell\nu$ shown in
Fig.~\ref{fig:fig14} (solid line) therefore strongly depends on the
$\Delta R$ cut 
imposed. For hadronic top decays, on the other hand, the milder increase
of the cross section ratio is mostly due to the logarithmic enhancement
factors associated with soft and/or collinear weak boson emission.

The ${\cal O}(\alpha)$ virtual weak corrections to $pp\to \bar tt$ for
on-shell top quarks were calculated in
Ref.~\cite{Bernreuther:2006vg}. For $p_T(t)=500$~GeV ($p_T(t)=1$~TeV) 
they reduce the LO $\bar tt$ cross section by about $5-6\%$ 
($10-11\%$), depending on the Higgs boson mass. Since top
quark decays and acceptance cuts equally affect 
the $\bar tt$ cross section with and without ${\cal O}(\alpha)$ virtual
weak corrections, these values can be used as an
estimate for how strongly the virtual weak corrections affect 
top quark pair production at the LHC. For $t\to bjj$ and
$p_T(t)=500$~GeV ($p_T(t)=1$~TeV), the $\bar ttV$ to $\bar tt$ cross
section is about 1\% (5\%), and 1.5\% (100\%) for $t\to b\ell\nu$.
If the top quark decays hadronically, weak boson emission effects thus
partially compensate the effect of the ${\cal O}(\alpha)$ virtual
weak corrections at  high $p_T(t)$. On the other hand, for a
semileptonically decaying top quark, weak boson emission effects
dominate over the ${\cal O}(\alpha)$ virtual corrections for
$p_T(t)>800$~GeV. However, only very few events are produced in this region
due to the separation cut (see Eq.~(\ref{eq:dr})) imposed.

Since the growth of the $\bar ttV$ to LO $\bar tt$ cross section ratio
in the $t\to b\ell\nu$ case is due to the kinematic cuts imposed, one
expects a similar effect if the weak 
boson is replaced by a jet. In other words, the QCD corrections to top
pair production at high $p_T(t)$ for $t\to b\ell\nu$ may be very large
if a $\Delta R$ cut is 
imposed on the final state particles and $pp\to\bar ttj$ may well
dominate the NLO QCD top pair cross section in this region. If this is
the case, the factorization and renormalization scale uncertainty of the
$\bar tt$ cross section in the high $p_T(t)$ region will be large, even
when NLO QCD corrections are taken into account. Before one can
determine whether it is important to take into account the ${\cal
O}(\alpha)$ weak corrections, it will therefore be necessary to
carefully investigate how NLO QCD corrections affect the $\bar tt$ cross
section at high top quark transverse momentum in the presence of
realistic acceptance cuts. 

The discussion presented here has focused on the transverse momentum
distribution of the top quark. Qualitatively similar results are found
for the $\bar tt$ invariant mass distribution.

\subsection{$t$-channel single top production}

Single top production provides an opportunity to study the $Wtb$
vertex~\cite{wtb}. To discriminate $t$-channel and $s$-channel single
top production, one makes use of the final state produced and the event
characteristics. In
$s$-channel single top production, the top (or anti-top) quark is
produced together with a high $p_T$ $b$-quark. Additional jets produced
via initial or final state radiation have a rapidity distribution peaked
in the central region, $|\eta|<2.5$. In $t$-channel single top
production, on the other hand, the $b$-quark in the final state
typically is soft, and the top quark is produced in association with a
light quark jet. The rapidity distribution of the light quark jet peaks at
$|\eta|\approx 3$. Jets originating from background processes such
as $\bar tt$ or $Wjj$ production, on the other hand, are predominantly
produced with rapidity $|\eta|<2.5$. To
select $t$-channel single top production, I therefore
require~\cite{cmstdr} one jet with
\begin{equation}
\label{eq:stop2}
p_T(j)>40~{\rm GeV}, \qquad 2.5<|\eta(j)|<4.5,
\end{equation}
and one $b$-jet (from top decay) with
\begin{equation}
\label{eq:bcut}
p_T(b)>35~{\rm GeV}, \qquad |\eta(b)|<2.5.
\end{equation}
Additional $b$-jets with $p_T(b)>35$~GeV and light quark or gluon jets
with $p_T(j)>40$~GeV are vetoed. The top quark
is identified through its semileptonic decay, with the following cuts
imposed on the charged lepton, $\ell=e,\,\mu$, and the missing
transverse momentum:
\begin{eqnarray}
\label{eq:stop}
p_T(\ell)>20~{\rm GeV,} &\qquad & |\eta(\ell)|<2.5, \\
\label{eq:ptm}
p\llap/_T>40~{\rm GeV.} &\qquad & 
\end{eqnarray}
I also assume that events do not contain a second charged lepton.
Finally, a 
\begin{equation}
\label{eq:stop3}
\Delta R(i,k)>0.4
\end{equation}
cut is imposed for $i,\, k=\ell,\,b,\,j$ and $i\neq k$. The cuts listed in
Eqs.~(\ref{eq:stop2}) --~(\ref{eq:stop3}) are similar to those used in
simulations by the CMS 
Collaboration~\cite{cmstdr}. Since events can only contain one $b$-jet,
$t$-channel single top production can be calculated treating the initial
state $b$-quark as a parton. Adopting this approach,
at lowest order (${\cal O}(\alpha^2)$), the process $pp\to tj$ and its
charge conjugate, $pp\to\bar tj$, contribute. 

The weak boson emission processes relevant for $t$-channel single top
production are $pp\to tjV$ and $pp\to\bar tjV$ which I collectively
denote as ``$tjV$ production''. $tjW$ 
production occurs at ${\cal O}(\alpha_s^2\alpha)$. The cross section for
the weak boson emission process thus is potentially as large as that
of the LO process. However, the central jet veto, and the
forward jet tagging requirement (see Eq.~(\ref{eq:stop2})), suppress the
$tjV$ cross section. Since events with more than one
charged lepton are vetoed, $tjV$ production with
$V=W\to\ell\nu$ and $V=Z\to\ell^+\ell^-$ 
does not contribute. In order to suppress events where
the $W$ boson from $t\to Wb$ decays hadronically and the other $W$
leptonically, a cut on the $b\ell p\llap/_T$ cluster transverse mass,
which peaks sharply at $m_t$, can be imposed. 

If the $W$ or $Z$ boson produced in association with the $tj$ system
decays hadronically, the jet satisfying Eq.~(\ref{eq:stop2}) may
originate from weak boson decay. This configuration can easily be taken
into account in the $Z$ case since the cross section for $tjZ$
production is finite. The ${\cal O}(\alpha_s^2\alpha)$ $tjW$ cross section, 
however, diverges for small jet transverse momenta, and a calculation of
$pp\to tW$, $W\to jj$ including NLO QCD corrections is
needed. The NLO QCD corrections for $tW$ production with $W\to\ell\nu$
have been calculated in Ref.~\cite{Campbell:2005bb} (see
also~\cite{zhu}) and were subsequently incorporated into {\tt MCFM}. In
order to estimate the NLO QCD $tW$, $W\to jj$ cross section, I
calculate $tW$, $W\to\ell\nu$ production, including NLO QCD corrections,
for the cuts specified in Eqs.~(\ref{eq:stop2}) --~(\ref{eq:stop3})
using {\tt MCFM}, and 
rescale the cross section to correct for the larger
$W\to jj$ decay rate. This approximation, of course, ignores QCD corrections
associated with the decay of the $W$ boson. 

The calculation of Ref.~\cite{Campbell:2005bb} also takes into account
the contributions from $\bar tt$ production where one of the $b$-quarks
is soft, and one of the light quark jets satisfies
Eq.~(\ref{eq:stop2}). Top pair production, where one of the $b$-quarks
is misidentified as a regular jet should then also be included in the
calculation. Assuming a probability of 40\% that a $b$-quark is
misidentified as a light quark or gluon jet, I find that this process
dominates weak boson emission for $p_T(t)\leq 200$~GeV. It drops very
rapidly at higher transverse momenta. In the intermediate region,
$p_T(t)=200-400$~GeV, $tW(j)$ and $tjZ$ production with $Z\to\bar\nu\nu$
are the main contributors, and for larger transverse momenta $pp\to
tjZ(\to\bar\nu\nu)$ dominates. 

\begin{figure}[t!] 
\begin{center}
\includegraphics[width=15.3cm]{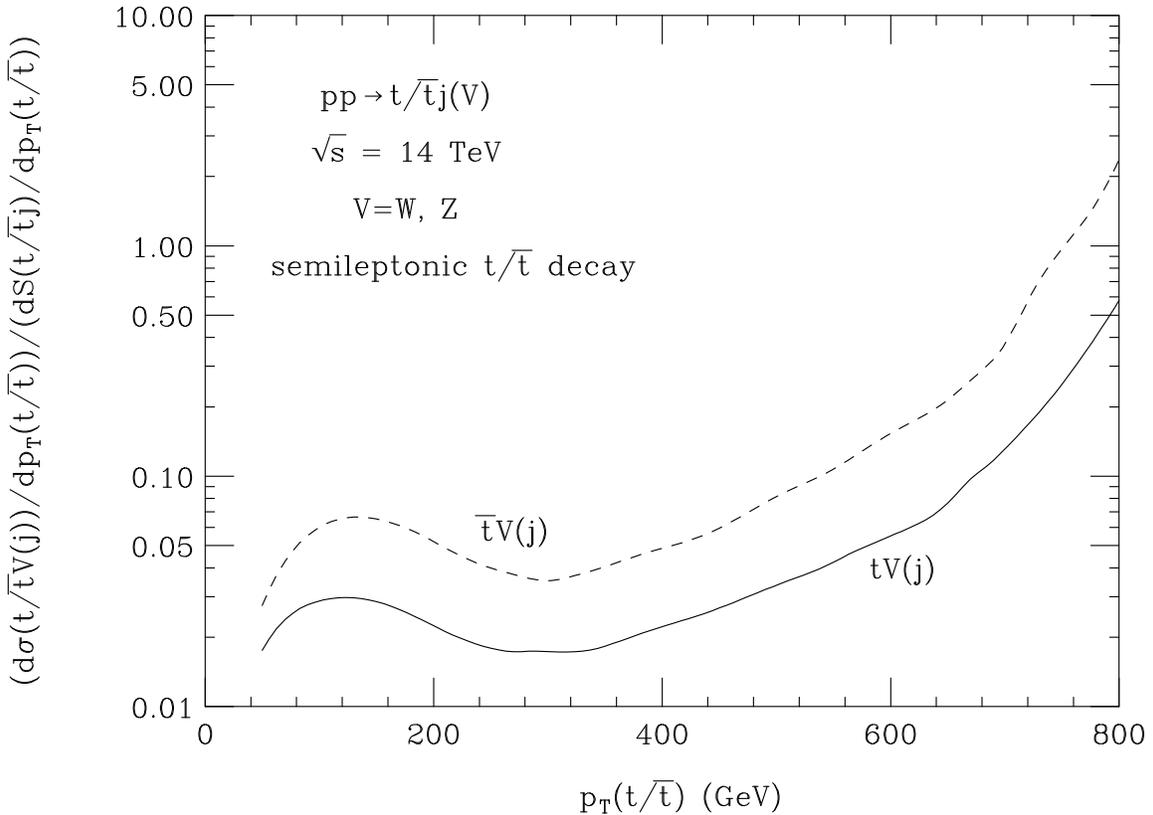}
\vspace*{2mm}
\caption[]{\label{fig:fig15} 
Ratio of the $tV(j)$ ($\bar tV(j)$) and the LO $tj$ ($\bar tj$) cross
section as a function of the transverse momentum of the $t$ ($\bar t$)
quark. The cuts imposed are discussed in the text.}
\vspace{-7mm}
\end{center}
\end{figure}

Since the jet is required to be in the forward rapidity region (see
Eq.~(\ref{eq:stop2})), the $p_T$ distribution of the top quark falls
very quickly. Vetoing additional jets in the event forces $tjV$ events
with $V\to jj$ into a phase space region which is very similar to that
of the LO $tj$ process. The $p_T(t)$ distribution for $pp\to tjV$ with
$V\to jj$ therefore also falls very steeply. However, there are no phase
space restrictions on the neutrinos produced in $pp\to tjZ$ with
$Z\to\bar\nu\nu$. This results in a much harder $p_T(t)$
distribution. As a result, $tjZ$ production with $Z\to\bar\nu\nu$ is the
dominating weak boson emission process at large top quark transverse
momenta. The $p_T(t)$ distribution for $pp\to tjZ$, $Z\to\bar\nu\nu$, is
also much harder than that of the LO $pp\to tj$ process. The $tjV$ to
LO $tj$ cross section ratio, which is shown in Fig.~\ref{fig:fig15},
therefore rises sharply at large $p_T(t)$.  
The dip at $p_T(t)\approx 300$~GeV is caused by the rapidly falling
cross section for $\bar tt$ production where the $b$-quark is
misidentified as a light quark jet. For 100~fb$^{-1}$, top quark
transverse momenta up to about 500~GeV should be accessible.

At large top quark transverse momenta, one expects that gluon radiation
frequently results in additional jets. The jet veto imposed suppresses
these effects and results in Sudakov form factors which may have a
significant impact on cross sections. However, the Sudakov form factors
are expected to partially cancel in the $tjV$ to $tj$ cross section
ratio. 

The combined ${\cal O}(\alpha)$ virtual weak and photonic radiative
corrections to $tj$ production were presented in
Ref.~\cite{Beccaria:2006ir} as a function of the parton center of mass
energy, $\sqrt{\hat s}$. For a top quark transverse momentum of
$p_T(t)=200$~GeV (400~GeV), the average parton center of mass
energy is $\sqrt{\hat s}\approx 1$~TeV (1.7~TeV). The combined virtual
weak and photonic ${\cal O}(\alpha)$ radiative corrections reduce the LO
$tj$ cross section by about 28\% (36\%) for $\sqrt{\hat s}=1$~TeV
(1.7~TeV). For comparison, the $tjV$ ($\bar tjV$) cross section is about
2\% (5\%) of the LO $tj$ ($\bar tj$) rate for both $p_T(t)=200$~GeV and
$p_T(t)=400$~GeV. Weak boson emission thus has   
a relatively small effect on the electroweak radiative corrections in
$t$-channel single top production at the LHC. Since the combined
theoretical~\cite{Sullivan:2004ie} and experimental~\cite{cmstdr} systematic
uncertainties on 
the $tj$ cross section are of ${\cal O}(10\%)$, it will be
important to take the ${\cal O}(\alpha)$ virtual weak radiative
corrections into account when 
analyzing $t$-channel single top production at the LHC. 

\subsection{$tW$ Production}

The $pp\to tW$ process contains two $W$ bosons and one $b$-quark in the
final state. The $W$ produced in association with the top quark has to
decay leptonically in order to be identified. The top quark may decay
semileptonically, $t\to\ell\nu b$ or hadronically, $t\to bjj$. Both
channels yield a signal of almost the same
significance~\cite{cmstdr}. Since it allows for a straightforward
reconstruction of the top quark transverse momentum, I only consider the
$t\to bjj$ final state here. Furthermore, since the cross sections for
$tW^-$ and $\bar tW^+$ production are equal, I focus on the process
$pp\to tW^-$.

To compute the $tW$ cross section,
I impose the cuts listed in Eqs.~(\ref{eq:bcut}) --~(\ref{eq:ptm}) on
the $b$-jet, the charged lepton and the missing transverse momentum. 
The non-$b$-like jets are required to have
\begin{equation}
\label{eq:tw}
p_T(j)>35~{\rm GeV}, \qquad |\eta(j)|<2.5.
\end{equation}
Lepton, $b$- and non-$b$-like jets are required to be isolated
in $\eta-\phi$ space by
\begin{equation}
\Delta R(i,k)>0.4
\end{equation}
($i,\, k=\ell,\,b,\,j$, $i\neq k$). The isolation cut strongly reduces
the $tW$ cross section at large top quark transverse momenta.
Finally, the invariant mass of the $bjj$ system
has to be within 20~GeV of the top quark mass:
\begin{equation}
\label{eq:twindow}
|m(bjj)-m_t|<20~{\rm GeV}.
\end{equation}

The dominant background to $tW$ production arises from $pp\to\bar
tt\to W^+W^-b\bar b$. To reduce the $\bar tt$ background, one tagged $b$-jet
and two non-tagged jets are required; events with additional $b$- or
non-tagged jets satisfying Eq.~(\ref{eq:tw}) are rejected. Likewise,
events are not allowed to have a second charged lepton. 

\begin{figure}[t!] 
\begin{center}
\includegraphics[width=15.3cm]{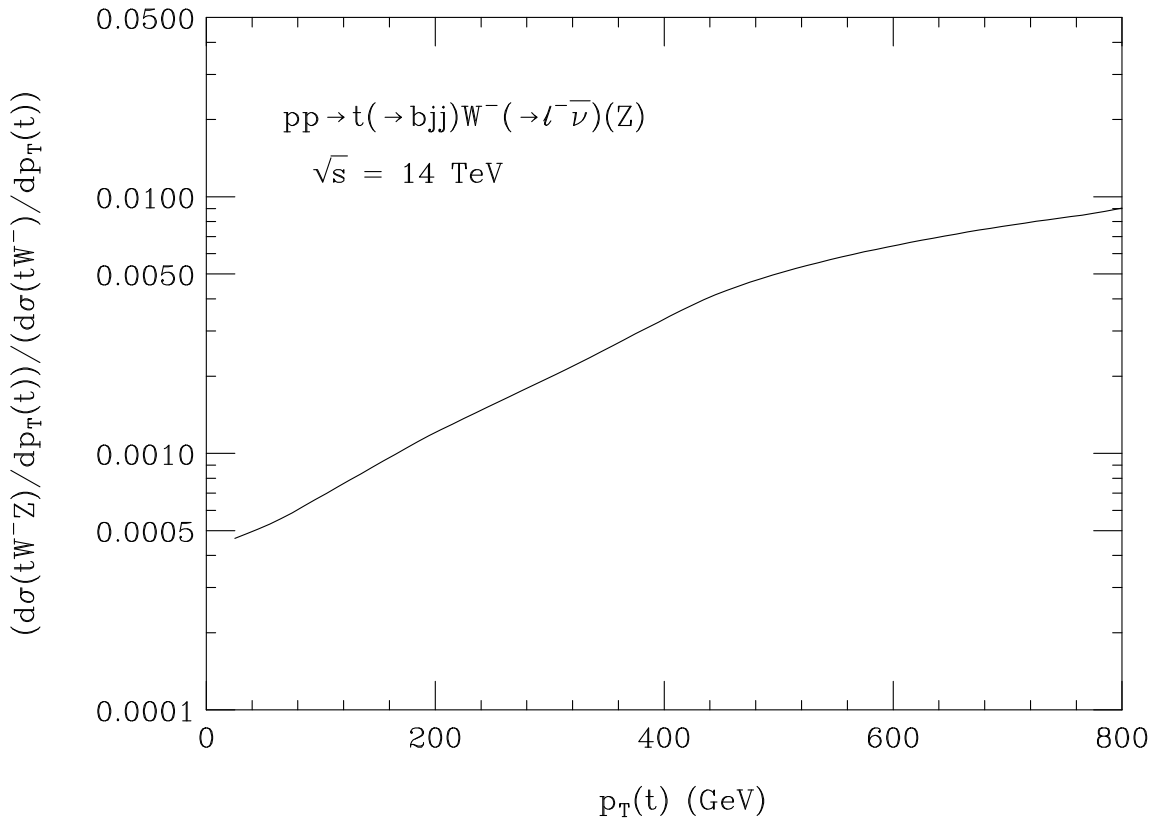}
\vspace*{2mm}
\caption[]{\label{fig:fig16} 
Ratio of the $tWZ$  and the LO $tW$ cross section as a function of the 
transverse momentum of the top quark. 
The cuts imposed are discussed in the text.}
\vspace{-7mm}
\end{center}
\end{figure}

The only weak boson emission process for $tW$ production is $pp\to
tWZ$. Since hadronic $Z$ decays are strongly suppressed by the jet veto
and Eq.~(\ref{eq:twindow}), it is not surprising that the ratio of the
$tWZ$ and LO $tW$ cross sections, which is shown in Fig.~\ref{fig:fig16}
as a function of $p_T(t)$, is small. 
Although the fraction of $tWZ$ events increases with $p_T(t)$, it is
below 1\% for top quark transverse momenta up to 800~GeV. For
100~fb$^{-1}$, top quarks with a transverse momentum of up to 600~GeV
will be produced in the $tW$, $t\to bjj$, $W\to\ell\nu$ channel with the
cuts listed in Eqs.~(\ref{eq:tw})~--~(\ref{eq:twindow}).

The combined ${\cal O}(\alpha)$ virtual weak and photonic radiative
corrections to $pp\to tW$ were presented in
Ref.~\cite{Beccaria:2006dt} as a function of the parton center of mass
energy, $\sqrt{\hat s}$. For a top quark transverse momentum of
$p_T(t)=200$~GeV, the average parton center of mass energy is $\sqrt{\hat
s}\approx 700$~GeV for which the combined virtual
weak and photonic ${\cal O}(\alpha)$ radiative corrections reduce the LO
$tW$ cross section by about 6\%.  For comparison, the $tWZ$ cross
section is about 0.13\% of the LO $tW$ rate for $p_T(t)=200$~GeV. Weak
boson emission thus is negligible in $tW$ production. The
electroweak ${\cal O}(\alpha)$ radiative corrections for $pp\to tW$ 
are substantially smaller than for $t$-channel single top production. 

The combined theoretical~\cite{nick} and experimental~\cite{cmstdr}
systematic uncertainties of the $tW$ cross section are about
$15-20\%$. They are considerably larger than the ${\cal O}(\alpha)$
electroweak radiative corrections.

\section{Summary and Conclusions}
\label{sec:sec6}

In the last five years, the ${\cal
O}(\alpha)$ electroweak radiative corrections for a number of processes
have been calculated. For some
processes~\cite{Kuhn:1999de,Kuhn:2005gv,Kuhn:2004em,Denner:2006jr},
higher order electroweak corrections were also calculated.
At high energies, the virtual weak corrections
become large and negative, due to soft and collinear logarithms of the
form $(\alpha/\pi)\log^2(\hat s/M_{W,Z}^2)$. In QCD and QED, the
corresponding terms diverge because gluons and photons are
massless. The divergencies cancel when real gluon and photon emission is
included in the calculation. Since the masses of the $W$ and $Z$ bosons
act as infrared regulators, there is no technical reason for including weak
boson emission in the calculation of weak radiative
corrections. Furthermore, since $W$ and $Z$ bosons decay, weak boson
emission leads to a different final state than the process
considered. $W$ and $Z$ boson emission 
therefore is ignored in most calculations of electroweak radiative
corrections. 

However, this does not mean that these contributions may not be important
at high energies, in particular in inclusive processes. In this paper, I
have investigated the importance of weak boson emission for those hadron
collider processes for which the ${\cal O}(\alpha)$ virtual weak
corrections are known to become large. In 
many cases, weak boson emission moderately reduces the effects of the
${\cal O}(\alpha)$ virtual weak corrections. Examples for processes
where this is the case 
are inclusive jet, isolated photon, $Z+1$~jet and Drell-Yan
production. In some processes, such as $W\gamma$ and $WZ$ production,
weak boson emission may become large, unless a jet veto is imposed and
the process becomes exclusive. 

Conclusions about the size of weak boson
emission effects may also depend on the observable considered. For
example in charged Drell-Yan production, $W$ and $Z$ radiation is much
more important in the lepton transverse momentum than in
the transverse mass distribution. An even more extreme case are the
$p_T$ and invariant mass distributions of the charged lepton pair in
$pp\to W^+W^-\to\ell_1^+\ell_2^- p\llap/_T$. In top pair production, the
acceptance cuts may significantly affect the relative importance of the
weak boson emission processes. Finally, in some processes
such as $s$-channel or $t$-channel single top production, the weak boson
emission processes involve gluon exchange, although the LO process is
purely weak. In this case, the cross section for the weak boson emission
processes is potentially much larger than that of the LO process.

The calculations presented in this paper demonstrate that it is not
possible to draw general conclusions about the importance of weak boson
emission. The 
relevant processes have to be calculated in each case. This is
straightforward and can be done efficiently using tools such as {\tt
MadEvent}. General purpose Monte Carlo programs such as {\tt
Pythia}~\cite{Sjostrand:2006za}, {\tt Herwig}~\cite{Corcella:2002jc} or
{\tt Sherpa}~\cite{Gleisberg:2003xi} do not take into account weak boson
emission. 

The purpose of this paper has been to investigate for which processes
and under what conditions weak boson emission may be important, not to
add weak boson emission to existing calculations of the ${\cal
O}(\alpha)$ virtual weak corrections. To do this, great care has to be
taken to use the same definitions and input parameters for both the
${\cal O}(\alpha)$ virtual weak corrections, and weak boson emission. I
have not done this; instead, for clarity, I opted for using one common
set of input parameters (see Eqs.~(\ref{eq:input1})
--~(\ref{eq:input2})). Furthermore, in QCD related processes, I have
taken into account QCD corrections to weak boson emission wherever
possible. Ultimately, for the analysis of LHC data, a tool
similar to {\tt MC@NLO}~\cite{Frixione:2006he} which contains the full
${\cal O}(\alpha)$ 
electroweak radiative corrections, including weak boson emission, for
all relevant processes, 
together with an interface to a general purpose Monte Carlo program, is
needed.

\acknowledgements
I would like to thank E.~Accomando, W.~Bardeen, J.~Campbell,
Y.~Gershtein, R.~Harris, J.~Huston, I.~Iashvili, N.~Kidonakis,
S.~Mrenna, S.~Pozzorini, P.~Skands, D.~Wackeroth and 
M.~Wobisch for useful discussions. I also would like to thank the
Fermilab Theory Group, where part of this work was carried out, for
its generous hospitality. This research was supported by the 
National Science Foundation under grant No.~PHY-0456681. 


\bibliographystyle{plain}

\end{document}